\documentclass[aps,pra,reprint,groupedaddress,longbibliography]{revtex4-1}

\usepackage{graphicx}
\usepackage{amssymb,amsmath}
\usepackage{bm}
\usepackage{dcolumn}
\usepackage[pdftex,breaklinks,colorlinks,citecolor=blue,linkcolor=blue,urlcolor=blue]{hyperref}
\usepackage{xcolor}

\usepackage{enumerate}
\newcommand{\mat}[1]{\bm{#1}} 

\begin{document}

\title{Multichannel quantum-defect theory for anisotropic interactions}

\author{Bo Gao}
\email[]{bo.gao@utoledo.edu}
\homepage[]{http://bgaowww.physics.utoledo.edu}
\affiliation{Department of Physics and Astronomy,
	Mail Stop 111,
	University of Toledo,
	Toledo, Ohio 43606,
	USA}

\date{August 18, 2020}

\begin{abstract}

We present a general formulation of multichannel quantum-defect theory (MQDT) for anisotropic long-range potentials. The theory unifies the treatment of atomic and molecular interactions of all types, and greatly expands the set of interactions that can be treated and understood systematically, including complex interactions involving molecules. In one exemplary manifestation, the theory provides a methodology to make the classification of atomic interactions based on the Periodic Table quantitative, instead of qualitative, and to generalize the Table to include molecular classes. Through the concept of effective potential, the theory further establishes a foundation for new classes of quantum theories for chemistry and for a broad range of quantum systems made of either a few or many atoms and/or molecules.

\end{abstract}


\maketitle

\section{Introduction}

Precision measurements enabled by cold atoms and especially cold molecules (see, e.g., Refs.~\cite{Ni08,Danzl08,Bala16,Bohn17}) have reaffirmed, unequivocally, many fundamental and humbling limitations in  our current understanding of nature. While we have the principles and equations of quantum mechanics, we are not yet able to predict many details of atom-molecule or molecule-molecule interactions, not even for some of the simplest systems for which we know every detail about their structures and internal modes \cite{Quemener05,Mayle12,Mayle13,Yang19,Gregory19}. When we learned that the atomic hypothesis -- \textit{``All things are made of atoms.''} -- was the most powerful and informative statement about nature \cite{Feynman63}, we did not anticipate running into difficulties at 3 or 4 atoms. Such limitations are disappointing and are in sharp contrast with our great success in understanding structures of all matters, from a single atom to much larger molecules such as DNA, cells, and materials of all kinds. 

The difficulty we encounter in understanding interaction is a reflection of a bigger contradiction in the current realization of the atomic picture (hypothesis).  While great progress has been made in understanding structures of all kinds, not only experimentally but also theoretically as exemplified by the density-functional theory (see, e.g., \cite{Burke12}), much less progress has been made in understanding their functionalities \cite{functionality}. We may know everything about the structure of a molecule, but we know little to nothing, theoretically, about how the same molecule interacts with others. Would it make a good drug? Would it serve as a good catalyst? We may know everything about the composition of a liquid, but we do not have a quantum theory of liquid for anything other than the liquid helium (see, e.g., Ref.~\cite{Leggett06}).

Why is there such a disparity regarding structure and functionality? How can the same atomic picture of the world be so powerful, yet at the same time so powerless? Understanding this disparity and overcoming the related difficulties are at the center of most scientific problems, from practical applications such as making chemistry and biophysics more ``physical'', to ``big questions'' such as those concerning the evolution of the universe. How and at what rate did the hydrogen molecules first form out of H atoms \cite{Turk11,Forrey13}? How did the hydrocarbons, water, and complex organic molecules first come about (see, e.g., Ref.~\cite{vanDishoeck17})? And how did life ultimately emerge spontaneously out of a collection of atoms? After all, everything is, or should be, in the atomic picture \cite{Feynman63}. 

This work is a part of our broader effort to reformulate the quantum realization of the atomic picture, to make it more useful for understanding functionalities. We briefly outline the rationale behind this broader effort to both emphasize and contextualize the central role of interactions, the focus of this work. Foundational concepts that serve the larger framework will be listed for clarity and for future reference.

We first examine the disparity in understanding structure and functionality. Its origin can be traced to the complexity of a many-body quantum system, specifically how the complexity depends on the types of interactions among its constituent particles \cite{Gao17a}. In the context of the atomic picture, we can state that
\begin{enumerate}[A.]
\it
\item Structure of matter is ``simple'' because it is fundamentally a many-electron problem. Functionalities are difficult because they are fundamentally $N$-atom problems.
\item An $N$-atom quantum system is much more difficult to understand, yet much more interesting than a many-electron system, primarily because atoms attract each other and can bind. This leads to the emergence of ``chemical complexity'' starting at $N=3$ atoms, accompanied by the ``arrangement'' concept. 
\end{enumerate}
Unlike most other ``emergences'' that occur in the thermodynamic limit of large numbers of atoms \cite{Anderson72}, the ``chemical complexity'', together with the ``arrangement'' concept, emerges at $N=3$. A preliminary look at the relationship between the complexity of a quantum system and the interaction among its constituent particles can be found in Ref.~\cite{Gao17a}. More discussions of this relationship, and more details of the concepts of arrangement and chemical complexity, will be presented in a separate publication \cite{Gao20d}. They will further explain and substantiate the insight and the wisdom in Feynman's more complete statement of the atomic picture: ``All things are made of atoms -- little particles that move around in perpetual motion, attracting each other when they are a little distance apart, but repelling upon being squeezed into one another.''\cite{Feynman63}, in which he recognized the importance of attraction. Indeed, without the attraction that is sufficiently strong to bind, a set of atoms in thermal equilibrium would already be in a maximum entropy state. No further ordering would have developed, and no human would have emerged to observe the world. The same attraction, however, also leads to so much complexity \cite{Quemener05,Mayle12,Mayle13,Gao20d} as to render our current quantum realization of the atomic picture, the one that we arrive at after the Born-Oppenheimer approximation (see, e.g., Ref.~\cite{Weinberg15}), mostly powerless quantitatively. 

The resolution of chemical complexity appears straightforward at first. As is common in a quantum theory, if it gets too difficult for investigation, we try to build a simpler, an effective, theory by focusing on a smaller range of energies. After all, the Born-Oppenheimer theory itself is already an effective theory resulting from eliminating most of the electronic degrees of freedom \cite{Weinberg15}, which are largely frozen at energies far below 1 eV or $10^4$ K. If we further note that it is the interaction that makes a many-body system difficult to analyze, it is not surprising that we construct a simpler effective theory by ``simplifying'' the potential, specifically by replacing the real potential with an effective potential that can nevertheless accurately describe interactions over a smaller range of energies. 

This method of building an effective theory through an effective potential is fundamental in quantum few-body and many-body physics \cite{HY57}. It is the foundation for existing theories of dilute quantum gases \cite{Leggett06,GPS08} and few-body theories of atoms \cite{Greene17}, in which the effective potential is often the Huang-Yang pseudopotential \cite{HY57}. For typical atoms and molecules with a long-range van der Waals $-C_6/r^6$ potential, such effective theories, commonly based on an $s$-wave pseudopotential \cite{HY57}, are applicable at temperatures and densities much lower than those determined by the corresponding van der Waals scales \cite{Gao09a}. Temperature-wise, it typically implies validity for 1 $\mu$K or lower temperatures. From such theories for dilute quantum gases and few-atom systems, we seem to be close to a general quantum theory of liquids, or a workable effective theory for chemical reactions. All we need is a better effective potential covering a greater range of energies and densities. It further seems that we can accomplish this within the pseudopotential approach \cite{HY57} by simply including more partial waves, until we realize that it is unrealistic because of the energy range needed and its implication on the number of partial waves required.

The traditional teaching of physics can give an incorrect impression that most phenomena at room temperature are classical in nature. A safer and more productive perspective to be instilled should instead be \textit{Almost everything interesting at room temperature is quantum in nature}. We give two examples to argue for this perspective. One is that the thermal de Broglie wavelength, $\lambda_T:=(2\pi\hbar^2/mk_B T)^{1/2}$ \cite{Huang87}, of a proton (or H atom) at 300 K is, oddly enough, 1 {\AA} to within 1\%. It is of the same order of magnitude as the interatomic spacing in a typical liquid or solid. Thus all hydrogen-rich condensed matters, including water and every life-related substances, are fundamentally quantum in nature even at room temperatures, since the wave property of H is intrinsically important. The importance of the concepts of hydrogen bond and pH in chemistry and biology are in this sense direct indicators of the importance of quantum effects in those fields even at room temperatures. (Similar arguments apply to other light elements including Li.) The second example, which is more directly related to our theory framework in its current stage, may be less obvious. It says that even in a gas phase, at a temperature where the de Broglie wavelength of an atom or molecule is much less than the interatomic spacing so that the motion between collisions is classical, the collision itself and the resulting interaction and/or reaction, is most likely quantum in nature, since
\begin{enumerate}[A.]
\it
\setcounter{enumi}{2}
\item unlike electrons, atoms and molecules are composite particles with internal degrees of freedom. Most of them or their aggregates have internal modes up to hundreds of kelvins, making quantum effects important even at those temperatures. Understanding functionalities in the real world thus requires quantum theories of interactions up to temperatures of hundreds of kelvins. 
\end{enumerate}
This point may be understood through an analogy. The internal modes provide the ``soft hands'' to capture a particle, which is often the first step towards whatever functionality of interest. The ``soft hands'' are physically scattering resonances which are intrinsically quantum in nature, whether they are Feshbach resonances \cite{Chin10} corresponding to virtual excitations of internal degrees of freedom, or shape resonances due to interference \cite{Gao08a}. The internal modes can be the rotational and vibrational modes of a molecule. They can also be the hyperfine and fine structures of an atom. It is the existence of these modes that makes atomic interactions interesting and keeps them from becoming classical far above the zero temperature. We note that even for a simplest atomic vapor such as an Ar vapor, with no (low-energy electronic) internal modes by itself, the formation of a single Ar$_2$ molecule would immediately introduce rovibrational modes, making its subsequent interaction with another Ar atom quantum. It is only above the boiling point of a substance, typically of the order of hundreds of kelvins, where atoms no longer aggregate and the kinetic energy is considerably greater than the typical fine structure splitting of 10 K, that the atomic interaction evolves towards classical. 

The requirement of an effective potential covering hundreds of kelvins is daunting. Hundreds-of-kelvins implies hundreds of contributing partial waves for typical atom-molecule and molecule-molecule interactions (see, e.g., Ref.~\cite{Gao13c,LYG14}). In a pseudopotential approach \cite{HY57}, even if one assumes one parameter per partial wave, the theory would have required too many parameters to qualify as a meaningful theory. And in reality, interaction in each partial wave is itself a complex and generally non-analytic function of energy (see, e.g., Ref.~\cite{LYG14}). This reasoning makes it clear that unless there exist universal behaviors in interaction that somehow relate different partial waves and energy variations, there can be no meaningful effective potential covering hundreds of kelvins. In other words, 
\begin{enumerate}[A.]
\it
\setcounter{enumi}{3}
\item the existence of meaningful theories for functionalities are predicated upon the existence of universal behaviors in atom-atom, atom-molecule, and molecule-molecule interactions over an energy range of hundreds of kelvins. 
\end{enumerate}

Progress on this front was made in connection with the quantum-defect theory (QDT) and multichannel quantum-defect theory (MQDT) for isotropic atomic interactions \cite{Mies84a,Burke98,Gao98b,Gao01,Gao05a,Gao08a,Ruzic13}, specifically the formulations that show the existences of universal behaviors \cite{Gao98b} and especially a partial-wave-insensitive formulation, which shows that different partial waves are related and can be described using the same set of parameters \cite{Gao01}. These advances led to the concept of effective potential for atom-atom interaction \cite{Gao03,Gao04a}, which was used to formulate universal behaviors for few-atom \cite{KG06} and many-atom quantum systems \cite{Gao04a,Gao05b} at higher densities and shorter length scales than those covered by the $s$-wave pseudopotential \cite{HY57}. In particular, Refs.~\cite{Gao04a,Gao05b} bridged the gap between a dilute Bose gas and the liquid helium, and predicted the existence of a gaseous Bose-Einstein condensate (BEC) branch for $^4$He \footnote{We still look forward to experimental realization and investigations of $^4$He gaseous BEC branch.}. These progresses made us believe, briefly, that we were close to a quantum formulation of the atomic picture. The remaining issues, however, turned out to be subtler and more difficult than we had anticipated at the time. Among the difficulties are proper representations of multichannel interactions and efficient progression to shorter length scales, in which progresses have gradually been made \cite{Gao09a,Gao11b,Gao16a,Hood20a}, though not yet incorporated in few-atom and many-atom theories. These difficulties are all minor when compared to the main obstacle which has been in the treatment of anisotropic potentials.

Most atom-atom interactions, and \textit{all} atom-molecule and molecule-molecule interactions, are intrinsically anisotropic even at long range \cite{Stone13} (see also Sec.~\ref{sec:prep}). The traditional MQDT, as pioneered by Seaton, Fano, and Greene \cite{Seaton83,Greene79,Greene82,Greene85}, and contributed to by many others, has been built mostly for isotropic long-range potentials with only very few exceptions \cite{Mittleman65,Clark79,Curik06,DKG09}. Without explicitly addressing anisotropic long-range potentials, such MQDT, despite its considerable success, would remain a specialized  theory, and cannot be a general theory of interactions nor a general theory of effective potential. 

It is in this context that this work takes a major step towards a general systematic understanding of interactions, and therein a new foundation for quantum realizations of the atomic picture. It shows that
\begin{enumerate}[A.]
\it
\setcounter{enumi}{4}
\item universal behaviors exist very generally for atomic interactions of all types, even those with anisotropic long-range potentials, and can be described very efficiently by a corresponding MQDT formulation. 
\end{enumerate}
One of the consequences of the theory is, simply put, to make the Periodical Table quantitative. For instance, the Periodical Table implies that atomic interactions can be grouped into types of group-$x$ with group-$y$ since all group-$x$ atoms behave similarly and so do all group-$y$ atoms. Our current explanation of this similarity, that all group-$x$-group-$y$ interactions share the same number of potential energy curves (PEC) with qualitative similarities, is however only qualitative. The new MQDT formulation will lead to quantitative and deeper understandings. Specifically,
\begin{enumerate}[A.]
\it
\setcounter{enumi}{5}
\item For each class of interactions, in the spirit of the Periodic Table and its generalizations, such as a group I atom or a $^2S$ atom with a $^1\Sigma$ molecule, MQDT for generally anisotropic long-range potentials (MQDTA) will provide a quantitative description of the interaction with a small number of short-range parameters and a few parameters characterizing the long-range potential.
\end{enumerate}
Different systems of the same class differ only in specific values of parameters. And the parameterization is in the very spirit of an effective theory:
\begin{enumerate}[A.]
\it
\setcounter{enumi}{6}
\item All parameters of the theory can be determined from experimental measurements even when they cannot be determined from \textit{ab initio} calculations.
\end{enumerate}
Through multiscale generalizations, the theory can be extended systematically to shorter length scales to yield theories that efficiently cover greater ranges of energies as needed. With such a systematic understanding of atomic and molecular interactions, we will finally be in a position to construct effective potentials and corresponding effective theories to better understand functionalities \cite{functionality}.

The rest of the paper is organized as follows. In Sec.~\ref{sec:prep}, we prepare for the MQDTA formulation through an overview of two-body interactions, including discussions of the conceptual foundation of MQDTA and the differences between isotropic and anisotropic potentials. A concise presentation of MQDTA follows in Sec.~\ref{sec:MQDTA}. Specifically, in Sec.~\ref{sec:qdtFuncs}, we define the QDT functions to be used in our formulations, including the reflection and transmission amplitudes associated with a generally anisotropic potential.  In Sec.~\ref{sec:Kmat}, a $K$-matrix formulation of interactions, for both scattering and bound state spectrum, is presented.  In Sec.~\ref{sec:Smat}, an $S$-matrix formulation using reflection and transmission amplitudes is presented. Section~\ref{sec:discussions} provides further discussions and clarifications, before we  conclude in Sec.~\ref{sec:conclusions}.

\section{Structure of two-body interactions}
\label{sec:prep}

Consider the interaction of two particles $A$ and $B$ in the absence of external fields. In the center-of-mass (COM) frame, it is described by a Hamiltonian
\begin{multline}
H = H_A+H_B-\frac{\hbar^2}{2\mu}\nabla_r^2+\hat{V} \\
	= H_A+H_B-\frac{\hbar^2}{2\mu}\frac{1}{r^2}\frac{\partial}{\partial r}\left(r^2\frac{\partial}{\partial r}\right) + \frac{1}{2\mu r^2}\hat{\bm{\ell}}^2\ +\hat{V}\;.
\label{eq:Ham}
\end{multline}
Here $H_A$ and $H_B$ are the Hamiltonians describing the internal degrees of freedom of particles $A$ and $B$, respectively. The $-\frac{\hbar^2}{2\mu}\nabla_r^2$ term describes their relative kinetic energy in the center-of-mass frame, with $\bm{r}:=\bm{r}_A-\bm{r}_B$ being the relative position vector between the center-of-mass's of the two particles and $\mu$ being the reduced mass. The operator $\hat{V}$ describes the interaction between the particles which satisfies
\[
\hat{V}\overset{r\to\infty}{\sim} 0 \;.
\]
The relative kinetic energy can be further split into a part associated with the relative radial motion and a part, specifically $\hat{\bm{\ell}}^2/2\mu r^2$, associated with the relative angular motion, with $\hat{\bm{\ell}}$ being the ``partial wave'' angular momentum operator associated with the direction of $\bm{r}$, to be labeled as $\hat{\bm{r}}$.

Expand the wave function as 
\begin{equation}
\psi_j = \sum_{i=1}^{N_\mathrm{ch}} \Phi_i F_{ij}(r)/r \;.
\label{eq:mchwfn}
\end{equation}
Here the channel functions $\{\Phi_i(r), i=1,2,\dots\}$ form an orthonormal basis for both the internal degrees of freedom of the particles and the angular part of their relative motion, with a parametric dependence on $r$. Specifically,
\[
\langle \Phi_i(r)|\Phi_j(r)\rangle = \delta_{ij} \;,
\]
where the inner product is over all degrees of freedom other than $r$. 
 $N_\mathrm{ch}$ is the total number of channels included in the expansion, and $j$ labels different solutions. Upon ignoring nonadiabatic couplings, the time-independent Schr\"{o}dinger equation at an energy $E$,
\[
H\psi = E\psi \;,
\]
can be rewritten, for a particular set of conserved quantities reflected in $\{\Phi_i(r)\}$, (and for the particular arrangement if rearrangement is possible),  as a set of coupled-channel (CC) equations for the radial part of the relative motion \cite{AD60}. They can be written in a matrix form as 
\begin{equation}
\left[-\frac{\hbar^2}{2\mu}\frac{d^2}{dr^2}\mat{1}+\frac{\hbar^2\bm{\ell}(\bm{\ell}+1)}{2\mu r^2}+\bm{V}(r)-\bm{\epsilon} \right]\mat{F} = 0 \;.
\label{eq:cceq}
\end{equation}
Here $\mat{F}$ is an $N_\mathrm{ch}\times N_\mathrm{ch}$ matrix made of elements $F_{ij}(r)$, with each column representing one linearly independent solution $\psi_j$ through Eq.~(\ref{eq:mchwfn}). $\mat{1}$ represents the unit matrix.  The $N_\mathrm{ch}\times N_\mathrm{ch}$ matrix $\bm{V}$ is the matrix representation of the interaction potential in the corresponding set of channel functions $\{\Phi_i(r)\}$, with elements
\[
V_{ij}(r) := \langle \Phi_i(r)|\hat{V}|\Phi_j(r)\rangle \;,
\]
$\mat{\epsilon}$ is the matrix representation of relative energies defined by
\[
\left(\mat{\epsilon}\right)_{ij} := E - \langle \Phi_i(r)| H_A+H_B |\Phi_j(r)\rangle \;,
\]
and we have used $\hbar^2\bm{\ell}(\bm{\ell}+1)$ to denote, very generally, the matrix representation of $\hat{\mat{\ell}}^2$, namely
\[
\left[\hbar^2\mat{\ell}(\mat{\ell}+1)\right]_{ij} := \langle \Phi_i(r)|\hat{\bm{\ell}}^2|\Phi_j(r)\rangle \;.
\]
While different representations corresponding to different choices of channel functions $\{\Phi_i(r)\}$ are generally possible, and are in fact very useful in efficient solutions and descriptions of interactions (see Sec.~\ref{sec:FT}), the most important representation for the definition of boundary conditions and scattering physical observables is the representation in the fragmentation channels. They correspond to channel functions that are, in the limit of $r\to\infty$, simultaneous eigenstates of $H_A$, $H_B$, and $\hat{\mat{\ell}}^2$. In the fragmentation channels, $\bm{\ell}$ is a diagonal matrix with elements $\ell_i$ being the partial-wave quantum number of channel $i$, and $\bm{\epsilon}$ is a diagonal matrix with elements $\epsilon_i=E-E_i$ being the energy relative to the channel energy $E_i:= E_{Ai}+E_{Bi}$. Unless otherwise stated, the set of CC equations in the fragmentation channels is our default and often our starting point for further discussions.

A two-body interaction in 3-D can always be formulated in this form. Prominent examples include its initial formulation for atom-molecule interaction by Arthurs and Dalgarno \cite{AD60}. Examples of formulations for atom-atom interaction including fine structures can be found in Refs.~\cite{Mies73,NU84,Zygelman94a}. A formulation of atom-atom interaction including hyperfine structures and nuclear statistics can be found in Ref.~\cite{Gao96}. Some early formulations for molecule-molecule interactions can be found in Refs.~\cite{Klar69,Green75}. A more recent formulation for atom-molecule and molecule-molecule interactions, by Tscherbul and Dalgarno, can be found in Ref.~\cite{Tscherbul10}.
The number of coupled channels, $N_\mathrm{ch}$, which is also the dimension of the matrix equation, Eq.~(\ref{eq:cceq}), is determined by the laws of conservation, specifically total angular momentum and parity conservations for interactions of electromagnetic origin, by the number of open channels at energy $E$, as characterized by $\epsilon_i>0$, and by the number of closed channels ($\epsilon_i<0$) required to achieve convergence.

Building an MQDTA as a general theory of interactions is possible because it need not rely on any specific characteristic of a particular system. Instead, it can be built based solely on the following two very general and very fundamental properties of an arbitrary two-body quantum system. 
\begin{enumerate}[(a)]
\it
\item The interaction potential at long range follows universal behaviors determined by the underlying fundamental interaction. (the $1/r$-property for atomic interactions)
\item The energy dependence of the interaction is determined primarily by the long-range potential, and reflected in the wave function at the long range. (the rigidity property)
\end{enumerate}
More specifically on property~(a), the electromagnetic nature of atomic interactions dictates that their corresponding potentials are real and have an asymptotic form of (see, e.g., \cite{Stone13})
\begin{equation}
\hat{V} \sim -\sum_{\{m\}}\frac{1}{r^m}\hat{C}_m,
\label{eq:LRVop}
\end{equation}
where the summation is over a set of positive integers $m$ determined by symmetry. We know this  as the multipole expansion in classical electrodynamics \cite{Jackson99}. The same structural form remains in the quantum theory (see, e.g., Ref.~\cite{Stone13}). We will call this the $1/r$-property of atomic interactions, which is basically a more specific statement on their electromagnetic origin.

The property~(b) is well known among practitioners of quantum mechanics, even outside the circle of QDT theories (see, e.g., Refs.~\cite{LeRoy70,vanKempen02}). In simple terms, it can be understood as that for every length scale in quantum mechanics, e.g., $r_0$, there is a corresponding energy scale, $(\hbar^2/2\mu)(1/r_0)^2$, which gets bigger as $r_0$ decreases. It implies that the short-range wave function is more rigid, more difficulty to change than the long-range wave function \cite{rigidity}. 
For atom-atom, atom-molecule and molecule-molecule interactions, this rigidity property, in combination with the same small electron-nucleus mass ratio ($m_e/M_N\ll 1$) that is behind the Born-Oppenheimer approximation \cite{Weinberg15}, leads also to the weak dependence of the short-range wave function on the partial wave, since a small $m_e/M_N$ makes the centrifugal energy term [$\hat{\bm{\ell}}^2/2\mu r^2$ in Eq.~(\ref{eq:Ham})] small compared to the interaction potential $\hat{V}$ at the short range \cite{Gao01}.

The combination of the $1/r$-property and the rigidity property provides the physical foundation for the existence of universal behaviors in interactions that MQDTA sets out to explore and represent. They imply that other than a few tightly-bound states and high-energy scattering states, all other quantum states, including both loosely-bound bound states and low-energy scattering states, follow universal behaviors determined by universal classes of long-range potentials. 

It is important to recognize that other than the monopole-monopole (the $m=1$ Coulombic) term, if present, all other terms of the potential in the long-range expansion, Eq.~(\ref{eq:LRVop}), are generally anisotropic. The only exception is when both particles have spherically symmetric charge distributions, namely when both are either structureless, such as an electron, or an atom (or ion) in an $S$ state. To put it more bluntly, all atom-molecule and molecule-molecule interactions are intrinsically anisotropic. Vast majority of atom-atom interactions, except when they are both in $S$ states, are also anisotropic. The prominence of the anisotropic long-range interaction highlights the importance of its treatment in a general theory of interactions.

The best-known example of isotropic atom-atom interactions is the one between two neutral atoms both in $S$ electronic states such as alkali-alkali interactions. It is described by a central potential
\[
V(r)\sim -C_6/r^6-C_8/r^8-C_{10}/r^{10} \;,
\]
where the $C_m$'s are constant van der Waals coefficients (see, e.g., \cite{Derevianko99,Derevianko01}).
All isotropic long-range potentials look similar, with each $1/r^m$ term characterized by a van der Waals coefficient describing the strength of the corresponding interaction. For any term with $m\neq 2$, the strength also defines a length scale $\beta_m := (2\mu C_n/\hbar^2)^{1/(m-2)}$ and a corresponding energy scale of $s_E^{(m)} := (\hbar^2/2\mu)(1/\beta_m)^2$ \cite{Gao08a}. The length scale corresponds to the radius at which the magnitude of $C_m/r^m$ is equal to a centrifugal energy of $\hbar^2/2\mu r^2$. The same magnitude also defines the corresponding energy scale $s_E^{(m)}$.

Anisotropic potentials, on the other hand, may look quite different from each other and generally require more parameters to characterize (see Appendix~\ref{sec:aniLRV}).  Anisotropy may show up either through an explicit dependence on $\hat{\bm{r}}$, or less explicitly through a coupling between different electronic states with different electronic angular momentum projections on the interparticle axis, or both. And different $1/r^m$ terms can have different anisotropy. Such diversity and complication have been some of the obstacles that have kept people, us included, from achieving a general MQDT for anisotropic potentials.

Fortunately, most of the differences and complexity in long-range potentials are superficial and there is a common structure behind them. In all cases, each term in Eq.~(\ref{eq:LRVop}) can still be characterized by a $C^{(x)}_m$ parameter measuring its overall strength and, if necessary, a few additional ``anisotropy parameters''. (See Appendix~\ref{sec:aniLRV}). And for any term with $m\neq 2$, this strength parameter, $C^{(x)}_m$, which we will still call an van der Waals coefficient, again defines a length scale
\begin{equation}
\beta_m^{(x)} := (2\mu C^{(x)}_m/\hbar^2)^{1/(m-2)} \;,
\end{equation}
and a corresponding energy scale
\begin{equation}
s_E^{(m,x)} := (\hbar^2/2\mu)(1/\beta_m^{(x)})^2 \;.
\end{equation}
Most importantly, regardless of any differences in details, the matrix representations of the potentials have a \textit{common structure} of 
\begin{equation}
\bm{V}(r)\sim -\sum_{\{m=n\}}^{n_X}\frac{1}{r^m}\bm{C}_m \;,
\label{eq:LRVms}
\end{equation}
for isotropic and all varieties of anisotropic potentials. An isotropic long-range potential has $\mat{C}_m$ matrices all diagonal in partial waves $\ell$, while an anisotropic long-range potential has at least one of the $\mat{C}_m$'s \textit{not} being diagonal in $\ell$. That is the only real difference. In writing Eq.~(\ref{eq:LRVms}), we have introduced $n_X$ to explicitly indicate the cutoff $n_X$ term in the expansion, and $n$ to indicate the dominant term, the term with the longest length scale and correspondingly the smallest energy scale. Thus the summation in Eq.~(\ref{eq:LRVms}) is in the order of decreasing length scales, which in most cases corresponds to increasing $m$ (e.g., when all terms are electric), but not always. For instance, a magnetic dipole-dipole term with $1/r^3$ radial dependence is in most cases much weaker, with a much shorter length scale, than the van der Waals $1/r^6$ dispersion term of electric origin. This $1/r^3$ term usually comes after the $1/r^6$ term despite having a smaller $m$ (see also Sec.~\ref{sec:GD}).

The common structures and properties of two-body interactions, as discussed above, enable a general formulation of MQDTA. Within a general framework for the long-range potential of the form of Eq.~(\ref{eq:LRVms}), the most important class of theories correspond to cases of a purely attractive potential with a single $n>2$ term, as in
\begin{equation}
\bm{V}(r)\sim -\frac{1}{r^n}\bm{C}_n \;,
\label{eq:LRVss} 
\end{equation}
with $n>2$. Here the requirement of a multichannel potential being purely attractive corresponds mathematically to $\mat{C}_n$ being positive-definite. This class of theories, to be presented in the rest of this paper, is the anisotropic equivalent of the isotropic theories of Refs.~\cite{Gao05a,Gao08a,Gao10b}. They cover most of neutral-neutral and charge-neutral atomic and molecular interactions in their ground electronic states, over typical temperature ranges of a few kelvins around a breakup threshold. From this fundamental class, other theories, including multiscale theories that cover broader ranges of energies, can be built (see Sec.~\ref{sec:GD}).

For consistency with the single-channel convention, we will call theories built for long-range potentials of a single $1/r^n$ term, as in Eq.~(\ref{eq:LRVss}), a single-scale theory, while theories for long-range potentials of multiple terms, as in Eq.~(\ref{eq:LRVms}), multiscale theories. This should be regarded as a convention that emphasizes different \textit{overall} length scales for different $1/r^m$ terms. It is not rigorous in the literal sense. As we will soon see in the next section, a single-term anisotropic potential has in fact multiple length scales, which can be understood as being associated with different interaction strength in different directions.

\section{MQDT for anisotropic long-range potentials (MQDTA)}
\label{sec:MQDTA}

We consider an $N_\mathrm{ch}$-channel two-body problem described by Eq.~(\ref{eq:cceq}) with potential $\mat{V}(r)$ behaving as that in Eq.~(\ref{eq:LRVss}). We will try to take advantage of the asymptotic behavior of the potential, $\mat{V}(r)\sim -\tfrac{1}{r^n}\bm{C}_n$ ($n>2$), while keeping in mind that it is still an $N_\mathrm{ch}$ channel problem, even at long range, with $N_\mathrm{ch}$ being potentially very large.

Recall from the standard scattering theory that $n>2$ implies that the potential $\mat{V}$ eventually goes away and we can match our solutions to a set of \textit{single-channel} free-particle solutions, specifically those of 
\begin{equation*}
\left[-\frac{\hbar^2}{2\mu}\frac{d^2}{dr^2}+\frac{\hbar^2\ell_i(\ell_i+1)}{2\mu r^2}
-\epsilon_i \right]v = 0 \;.
\label{eq:free}
\end{equation*}
This characteristic underlies the definitions of the scattering $K$ matrix and $S$ matrix.

The easiest improvement upon the standard theory is that, instead of matching to free-particle solutions, we match to a set of single-channel solutions for
\begin{equation}
\left[-\frac{\hbar^2}{2\mu}\frac{d^2}{dr^2}+\frac{\hbar^2\ell_i(\ell_i+1)}{2\mu r^2}
-\frac{(\mat{C}_n)_{ii}}{r^n}-\epsilon_i \right]v = 0 \;.
\label{eq:scLRss} 
\end{equation}
In other words, we ignore the off-diagonal coupling in $\bm{V}(r)\sim -\tfrac{1}{r^n}\bm{C}_n$, while keeping the diagonal terms. This is, in essence, what we do currently in applying the MQDT developed for isotropic long-range potentials to problems with anisotropic long-range potentials, specifically in atom-molecule and molecule-molecule interactions \cite{Gao10b,Croft11,Mayle12,Croft12,Mayle13,Croft13,Hazra14a,Hazra14b}. This approach is valid mathematically since if $\mat{V}$, in its entirety, goes away at a sufficiently large radius $r$, it is certainly fine if we ``only'' ignore the off-diagonal coupling at sufficiently large $r$. It never works any worse than the standard scattering theory. Furthermore, since the energy differences between different channels, $\Delta E$, when they are not degenerate, are always much greater than the energy scale associated with the (dominant) long-range interaction, $s_E^{(n,x)}$,  it is in fact a very good approximation, in the absence of external fields, to ignore channel coupling, provided the channels involved have different threshold energies. To be more precise, in the absence of external fields, $\Delta E$ is at least of the order of a hyperfine splitting, $\Delta E^{\mathrm{hf}}$, with a typical magnitude of  0.1 K. It is much greater than the typical $s_E^{(n,x)}$ such as 1 mK or less for the $1/r^6$ van der Waals potential \cite{Gao09a,Gao10a}. The coupling between non-degenerate states only becomes important for $|C^{(x)}_n/r^n|>\sim \Delta E^{\mathrm{hf}}$, corresponding to $r\ll \beta^{(x)}_n$. It means that the coupling is important only at the short range, not in the long-range QDT region of $r\sim \beta^{(x)}_n$ or greater $r$.

The limitation of isotropic MQDT arises when there are degenerate channels that are coupled by the long-range potential, which is precisely one of the effects of an anisotropic potential. For most thresholds, each corresponding to an $E_i  = E_{Ai}+E_{Bi}$, there are generally multiple degenerate channels that differ only in partial wave $\ell$ and are coupled by the long-range anisotropic potential. In those cases, the long-range solutions need to incorporate the long-range coupling explicitly. In other words, instead of single channel solutions of Eq.~(\ref{eq:scLRss}), the long-range solutions should be those for
\begin{equation}
\left[-\frac{\hbar^2}{2\mu}\frac{d^2}{dr^2}\mat{1}+\frac{\hbar^2\mat{\ell}_d(\mat{\ell}_d+\mat{1})}{2\mu r^2}
-\frac{1}{r^n}(\mat{C}_n)_{dd}-\epsilon_d\mat{1} \right]\mat{v} = 0 \;.
\label{eq:cceqLRss}
\end{equation}
It is an equation of $N_d\times N_d$ dimension for a degenerate manifold, labeled by $d$, of $N_d$ channels all having the same channel energy $E_d := E_i, (i\in d)$ and therefore all the same relative energy, $\epsilon_d :=E-E_d$. Here $(\mat{C}_n)_{dd}$ is the $N_d\times N_d$ principle submatrix of $\mat{C}_n$ representing the long-range interaction within the degenerate manifold $d$, and is in general real and symmetric, but not necessarily diagonal. 

The distinction of MQDTA is that it is built upon solutions of Eq.~(\ref{eq:cceqLRss}) instead of single-channel solutions of Eq.~(\ref{eq:scLRss}). In this way we ensure that the ``short-range'' parameters in the theory are truly short-range in nature, thus have the weak energy and partial-wave dependences, which are essential for an effective theory. More completely, we build MQDTA by separating the $N_\mathrm{ch}$-channel long-range problem into a set of degenerate channels of much smaller dimensions $N_d$, satisfying $\sum_d N_d = N_\mathrm{ch}$, with long-range coupling within the degenerate manifolds fully accounted for in the long-range solutions.

\subsection{QDT functions for anisotropic long-range potentials}
\label{sec:qdtFuncs}

QDT functions are what we use to characterize long-range solutions. For isotropic potentials, they are associated with solutions of single-channel equations, Eq.~(\ref{eq:scLRss}) \cite{Gao08a}. For anisotropic potentials, they are associated with solutions of Eq.~(\ref{eq:cceqLRss}) for a set of degenerate manifolds $\{d=1,2,\dots\}$. Since the equations are of the same form for all $d$, we drop the index $d$ for simpler notation. Thus solutions of
\begin{equation}
\left[-\frac{\hbar^2}{2\mu}\frac{d^2}{dr^2}\mat{1}+\frac{\hbar^2\mat{\ell}(\mat{\ell}+1)}{2\mu r^2}
-\frac{1}{r^n}\mat{C}_n-\epsilon\mat{1} \right]\mat{v} = 0 \;,
\label{eq:aniLRss} 
\end{equation}
with $\mat{C}_n$ being an $N_d\times N_d$ real and symmetric matrix, will define the QDT (matrix) functions for MQDTA. As mentioned earlier, we limit ourselves here to cases of purely attractive potentials corresponding to $\mat{C}_n$ being positive definite \cite{positiveDefinite}. 

Equation~(\ref{eq:aniLRss}) is, in general, itself a CC equation, except that it has a much smaller dimension than the original, and all coupled channels are degenerate in energy. It is similar in form to the single-channel equation [cf. Eq.~(\ref{eq:scLRss})] that defines the single-channel QDT \cite{Gao08a} and the MQDT for isotropic potentials \cite{Gao05a,Gao10b}. The only difference is that Eq.~(\ref{eq:aniLRss}) is a matrix equation with $\mat{C}_n$ generally non-diagonal. The similarity means that Eq.~(\ref{eq:aniLRss}) has many of the same qualitative properties as the corresponding single-channel equation \cite{Gao08a}. In particular, the $1/r^n$ ($n>2$) interaction term dominates at small $r$, while the $1/r^2$ centrifugal term dominates at large $r$. The transition, from the behavior determined by the essential singularity at the origin, to the behavior determined by the essential singularity at $r=\infty$, is the most important characteristic of the QDT equation \cite{Gao08a}. This transition occurs around $r\sim\beta_n^{(x)}$.

Defining a scaled radius as $r_s := r/\beta_n^{(x)}$, and a scaled energy as $\epsilon_s := \epsilon/s_E^{(n,x)}$, Eq.~(\ref{eq:aniLRss}) can be written in a dimensionless form as
\begin{equation}
\left[\frac{d^2}{dr_s^2}\mat{1}-\frac{\mat{\ell}(\mat{\ell}+\mat{1})}{r_s^2}
+\frac{1}{r_s^n}\bar{\mat{C}}_n+\epsilon_s\mat{1} \right]\mat{v} = 0 \;,
\label{eq:aniLRsss} 
\end{equation}
where $\bar{\mat{C}}_n := \mat{C}_n/C^{(x)}_n$ is a scaled and dimensionless $\mat{C}_n$, with elements of the order of 1. In the isotropic case, $\bar{\mat{C}}_n=\mat{1}$, and the scaled equation would tell us that the QDT functions are universal functions of the scaled energy $\epsilon_s$ and the partial wave $\ell$ \cite{Gao08a}. They depend on energy only through $\epsilon_s$, and the wave functions can be defined to depend on $r$ only through $r_s$. In anisotropic cases, Eq.~(\ref{eq:aniLRsss}) tells us that they are similar universal functions, with possible additional parametric dependences on additional anisotropy parameters required to characterize $\hat{C}_n$ (see Appendix~\ref{sec:aniLRV}).

Another important general property of the QDT equation, Eq.~(\ref{eq:aniLRss}) or its scaled version Eq.~(\ref{eq:aniLRsss}), is that it has $2N_d$ linearly independent solutions. Let $\mat{f}$ and $\mat{g}$ be two $N_d\times N_d$ matrices whose columns, together, represent one such set of solutions. Their Wronskian, defined by
\begin{equation}
\mat{W}_{r_s}(\mat{f},\mat{g}) := \mat{f}^T\left(\frac{d}{d r_s}\mat{g}\right)
	-\left(\frac{d}{d r_s}\mat{f}^T\right)\mat{g} \;,
\end{equation}
with superscript $T$ denoting matrix transpose, is a constant matrix in the sense of 
\[
\frac{d}{dr_s}\mat{W}_{r_s} = \mat{0} \;.
\]
This constancy of Wronskian is easily verified by substitution. Note that we have included the derivative variable $r_s$ explicitly in our Wronskian notation to distinguish different, but equally valid, definitions such as $\mat{W}_r(\mat{f},\mat{g})$, which is defined with respect to $r$ instead of $r_s$.

\subsubsection{The base pair of reference functions and the real propagation matrices}

There is considerable freedom in picking reference functions, specific sets of solutions for Eq.~(\ref{eq:aniLRss}) or (\ref{eq:aniLRsss}), for the MQDTA formulation. One consideration is that at least one of the sets should be defined with partial-wave-independent boundary conditions at the short range, specifically in the limit of $r\to 0$, to better enable a partial-wave-insensitive formulation when the underlying interaction has such characteristics \cite{Gao01}. For our formulation, we choose such a pair, which we call the base pair $(\mat{f}^c, \mat{g}^c)$, to be defined such that they automatically reduce to their isotropic counterparts in cases of isotropic potentials \cite{Gao08a}.

Specifically, at sufficiently small $r\ll \beta_n^{(x)}$, the $1/r^2$ term is negligible compared to the $1/r^n$ ($n>2$) term, and Eq.~(\ref{eq:aniLRss}) becomes.
\begin{equation*}
\left[-\frac{\hbar^2}{2\mu}\frac{d^2}{dr^2}\mat{1}-\frac{1}{r^n}\mat{C}_{n} \right]\mat{v} = 0 \;.
\end{equation*}
This matrix equation can be diagonalized by diagonalizing $\mat{C}_n$, specifically by going to its eigenbasis defined by
\begin{equation}
\mat{C}_n|\alpha\rangle = C_{n\alpha}|\alpha\rangle \;,
\end{equation}
where $C_{n\alpha}, \alpha = 1, 2, \cdots, N_d$ are all positive for a purely attractive potential. This equation defines what we call the $\mat{C}_{nd}$-basis or the van der Waals basis. Each eigenvalue $C_{n\alpha}$ has a corresponding length scale
\begin{equation}
\beta_{n\alpha} := (2\mu C_{n\alpha}/\hbar^2)^{1/(n-2)} \;,
\end{equation}
implying, as we mentioned earlier, that an anisotropic potential, even with a single $-\mat{C}_n/r^n$ term, has, rigorously speaking, multiple scales.

The base pair of reference functions $\mat{f}^c$ and $\mat{g}^c$, both $N_d\times N_d$ matrices, are defined such that they are diagonal in the $\mat{C}_{nd}$-basis with asymptotic behavior at small $r$ given by 
\begin{subequations}
\begin{align}
\widetilde{\mat{f}}^c & := \mat{U}_{nd}^T\mat{f}^c \mat{U}_{nd} \;, \\
	& \overset{r\to 0}{\sim} 
	(2\mat{\beta}_{ns}/\pi)^{1/2}\mat{r}_s^{n/4}
	\cos\left(\mat{y}-\pi/4 \right) \;, 
\label{eq:fctasy0}\\
\widetilde{\mat{g}}^c & := \mat{U}_{nd}^T\mat{g}^c \mat{U}_{nd} \;, \\
	&\stackrel{r\to 0}{\sim}
	-(2\mat{\beta}_{ns}/\pi)^{1/2}\mat{r}_s^{n/4}
	\sin\left(\mat{y} -\pi/4 \right) \;,
\label{eq:gctasy0}
\end{align}
\end{subequations}
for all energies. Here $\mat{\beta}_{ns}$, $\mat{r}_s$, and $\mat{y}$ are diagonal matrices in the $\mat{C}_{nd}$-basis.
\[
\mat{\beta}_{ns} := \mat{\beta}_n/\beta_n^{(x)} \;,
\]
where $\mat{\beta}_n$ is a diagonal matrix with elements $\beta_{n\alpha}$. And we have defined
\[
\mat{r}_s := r/\mat{\beta}_n = r_s/\mat{\beta}_{ns} \;,
\]
and
\[
\mat{y} := \frac{2}{(n-2)}\mat{r}_s^{-(n-2)/2} \;.
\]
More explicitly in the fragmentation channels, $\mat{f}^c$ and $\mat{g}^c$ are solutions of Eq.~(\ref{eq:aniLRss}) or (\ref{eq:aniLRsss}) with asymptotic behaviors of 
\begin{subequations}
\begin{align}
\mat{f}^c & = \mat{U}_{nd}\widetilde{\mat{f}}^c \mat{U}_{nd}^T \;, \\
	& \overset{r\to 0}{\sim} 
	\mat{U}_{nd} (2\mat{\beta}_{ns}/\pi)^{1/2}\mat{r}_s^{n/4}
	\cos\left(\mat{y}-\pi/4 \right) \mat{U}_{nd}^T \;, 
\label{eq:fcasy0}\\
\mat{g}^c & = \mat{U}_{nd}\widetilde{\mat{g}}^c \mat{U}_{nd}^T \;, \\
	&\stackrel{r\to 0}{\sim}
	- \mat{U}_{nd} (2\mat{\beta}_{ns}/\pi)^{1/2}\mat{r}_s^{n/4}
	\sin\left(\mat{y} -\pi/4 \right) \mat{U}_{nd}^T \;,
\label{eq:gcasy0}
\end{align}
\end{subequations}
where $\mat{U}_{nd}$, with elements $(\mat{U}_{nd})_{i\alpha} = \langle i|\alpha\rangle$, is the matrix of orthogonal transformation between the fragmentation channels and the $\mat{C}_{nd}$-basis channels.
The normalization constants are chosen such that their Wronskian matrix, with respect to $r_s$, is given by
\begin{equation}
\mat{W}_{r_s}(\mat{f}^c,\mat{g}^c) = \frac{\pi}{2}\mat{1}\;,
\label{eq:Wfcgc}
\end{equation}
with $\mat{1}$ again being the unit matrix.

The base pair of solutions $\mat{f}^c$ and $\mat{g}^c$ are fully defined by Eq.~(\ref{eq:aniLRss}) or (\ref{eq:aniLRsss}) and their asymptotic behaviors as specified by Eqs.~(\ref{eq:fcasy0}) and (\ref{eq:gcasy0}). Since $n>2$ and the particles are asymptotically free, they have, for $\epsilon>0$, large-$r$ asymptotic behaviors given by 
\begin{subequations}
\begin{align}
\mat{f}^{c} &\overset{r\to \infty}{\sim}
	\sqrt{\frac{2}{\pi k_s}}\left[\sin\left(k_s r_s\mat{1}
	-\frac{1}{2}\mat{\ell}\pi\right) \mat{Z}^{c}_{sf} \right. \nonumber\\
	&- \left. \cos\left(k_s r_s\mat{1}-\frac{1}{2}\mat{\ell}\pi\right)\mat{Z}^{c}_{cf}\right] \;,
\label{eq:fcLRpe}\\	 
\mat{g}^{c} &\overset{r\to \infty}{\sim}
	\sqrt{\frac{2}{\pi k_s}}\left[\sin\left(k_s r_s\mat{1}
	-\frac{1}{2}\mat{\ell}\pi\right) \mat{Z}^{c}_{sg} \right.\nonumber \\
	&- \left.\cos\left(k_s r_s\mat{1}-\frac{1}{2}\mat{\ell}\pi\right) \mat{Z}^{c}_{cg}\right] \;,
\label{eq:gcLRpe}
\end{align}
\end{subequations}
where $k_s := k\beta_n^{(x)}, k :=(2\mu\epsilon/\hbar^2)^{1/2}$, and therefore $k_s r_s = k r$. This behavior defines 4 $N_d\times N_d$ $\mat{Z}^{c}_{xy}$ matrices, which together can be grouped into a single $2N_d\times 2N_d$ $\mat{Z}^c$ matrix,
\begin{equation}
\mat{Z}^c :=
\begin{pmatrix}
\mat{Z}^c_{sf} & \mat{Z}^c_{sg} \\
\mat{Z}^c_{cf} & \mat{Z}^c_{cg}
\end{pmatrix}
\;.
\end{equation}
For $\epsilon<0$, the large-$r$ asymptotic behaviors of the QDT base pair define a $\mat{W}^{c}$ matrix function, as in
\begin{subequations}
\begin{align}
\mat{f}^{c} &\overset{r\to \infty}{\sim}
	\frac{1}{\sqrt{\pi\kappa_s}}\left[ e^{-\kappa_s r_s\mat{1}}\mat{W}^{c}_{+f}
	+e^{+\kappa_s r_s\mat{1}}\mat{W}^{c}_{-f}\right] \;,
\label{eq:fcLRne} \\
\mat{g}^{c} &\overset{r\to \infty}{\sim}
	\frac{1}{\sqrt{\pi\kappa_s}}\left[ e^{-\kappa_s r_s\mat{1}}\mat{W}^{c}_{+g}
	+e^{+\kappa_s r_s\mat{1}}\mat{W}^{c}_{-g}\right] \;,
\label{eq:gcLRne}
\end{align}
\end{subequations}
where $\kappa_s := \kappa\beta_n^{(x)}, \kappa :=(-2\mu\epsilon/\hbar^2)^{1/2}$, and therefore $\kappa_s r_s = \kappa r$. This behavior defines  4 $N_d\times N_d$ $\mat{W}^{c}_{xy}$ matrices, which together can be grouped into a single $2N_d\times 2N_d$ $\mat{W}^c$ matrix
\begin{equation}
\mat{W}^c :=
\begin{pmatrix}
\mat{W}^c_{+f} & \mat{W}^c_{+g} \\
\mat{W}^c_{-f} & \mat{W}^c_{-g}
\end{pmatrix}
\;.
\end{equation}
The Wronkian relation of Eq.~(\ref{eq:Wfcgc}) implies that not all submatrices of $\mat{Z}^c$ and $\mat{W}^c$ are independent, but are related by
\begin{equation}
(\mat{Z}^c_{sf})^T\mat{Z}^c_{cg}-(\mat{Z}^c_{cf})^T\mat{Z}^c_{sg}=\mat{1} \;,
\end{equation}
and
\begin{equation}
(\mat{W}^c_{+f})^T\mat{W}^c_{-g}-(\mat{W}^c_{-f})^T\mat{Z}^c_{+g}=\mat{1} \;.
\end{equation}
The $\mat{Z}^c$ and $\mat{W}^c$ matrices describe, in the standing wave representation, the propagation through the long-range $-\mat{C}_n/r^n$ potential from $r\ll\beta_n^{(x)}$ to $r\gg\beta_n^{(x)}$, for $\epsilon>0$ and $\epsilon<0$, respectively. Given a linear superposition of $\mat{f}^c$ and $\mat{g}^c$, we can obtain from $\mat{Z}^c$ and $\mat{W}^c$ the asymptotic behavior of the wave function in the limit of $r\to\infty$ where both the scattering boundary condition and the boundary condition for bound states are defined. This physical picture corresponds to the $K$-matrix formulation of MQDTA to be carried out in detail in Sec.~\ref{sec:Kmat}.

Another independent and important interpretation of the $\mat{Z}^c$ and $\mat{W}^c$ matrices is that they relate the base pair $(\mat{f}^c,\mat{g}^c)$ to other useful pairs of reference functions, in particular the $(\mat{s},\mat{c})$ pair that is directly related to the definition of the physical $K$ matrix, and the $(\mat{f}^{o+},\mat{f}^{o-})$ pair that is important both for the definition of the $S$ matrix and for the definitions of reflection and transmission amplitudes.

Specifically, at sufficiently large $r\gg\beta_n^{(x)}$ where the  $1/r^{n}$ ($n>2$) term is small compared to $1/r^2$ term, the solutions of Eq.~(\ref{eq:aniLRss}) are also free-particle solutions satisfying
\begin{equation*}
\left[-\frac{\hbar^2}{2\mu}\frac{d^2}{dr^2}\mat{1}+\frac{\hbar^2\mat{\ell}(\mat{\ell}+1)}{2\mu r^2}-\epsilon\mat{1} \right] v = 0 \;.
\end{equation*}
For $\epsilon>0$, we define a pair of solutions of Eq.~(\ref{eq:aniLRss}) or (\ref{eq:aniLRsss}), the $(\mat{s},\mat{c})$ pair, with large-$r$ asymptotic behaviors of
\begin{subequations}
\begin{align}
(\mat{s})_{ij} &\overset{r_s\gg 1}{\sim} \delta_{ij}\sqrt{\frac{2}{\pi k_s}}\, (k_s r_s)\, j_{\ell_i}(k_s r_s) \;,\\
&\overset{r\to \infty}{\sim} \delta_{ij}\sqrt{\frac{2}{\pi k_s}}\, \sin\left(k_s r_s-\frac{1}{2}\ell_i\pi\right) \;,\\
(\mat{c})_{ij} &\overset{r_s\gg 1}{\sim} \delta_{ij}\sqrt{\frac{2}{\pi k_s}}\, (k_s r_s)\, y_{\ell_i}(k_s r_s) \;,\\
&\overset{r\to \infty}{\sim} -\delta_{ij}\sqrt{\frac{2}{\pi k_s}}\, \cos\left(k_s r_s-\frac{1}{2}\ell_i\pi\right) \;.
\end{align}
\end{subequations}
They are normalized such that
\[
\mat{W}_{r_s}(\mat{s},\mat{c}) = \frac{\pi}{2}\mat{1}\;.
\label{eq:Wsc}
\]
Different pairs of solutions of the same linear equation are related by linear superpositions of constant coefficients, which can be determined in any region in space. Comparing the $(\mat{s}, \mat{c})$ pair and the $(\mat{f}^c, \mat{g}^c)$ pair in the region of $r\to\infty$, through Eqs.~(\ref{eq:fcLRpe}) and (\ref{eq:gcLRpe}), we have
\begin{align*}
\mat{f}^{c} &=
	\mat{s}\mat{Z}^{c}_{sf} + \mat{c}\mat{Z}^{c}_{cf} \;, \\
\mat{g}^{c} &=
	\mat{s}\mat{Z}^{c}_{sg} + \mat{c}\mat{Z}^{c}_{cg} \;,
\end{align*}
or in a more concise matrix form as
\begin{equation}
\begin{pmatrix}
\mat{f}^c & \mat{g}^c
\end{pmatrix}
=
\begin{pmatrix}
\mat{s} & \mat{c}
\end{pmatrix}
\mat{Z}^c
=
\begin{pmatrix}
\mat{s} & \mat{c}
\end{pmatrix}
\begin{pmatrix}
\mat{Z}^c_{sf} & \mat{Z}^c_{sg} \\
\mat{Z}^c_{cf} & \mat{Z}^c_{cg}
\end{pmatrix}
\;.
\label{eq:cscpe}
\end{equation}
Thus the $\mat{Z}^c$ matrix can also be understood as the matrix that relates the $(\mat{f}^c, \mat{g}^c)$ pair and the $(\mat{s}, \mat{c})$ pair at positive energies.

The $\mat{W}^{c}$ matrix function can be similarly understood in terms of another pair of reference functions, $(\mat{f}^{o+},\mat{f}^{o-})$, corresponding to the outgoing and incoming waves in the outer region of $r\gg \beta_n^{(x)}$, respectively. They are defined for positive energies by
\begin{subequations}
\begin{align}
\mat{f}^{o+} &\overset{r\to\infty}{\sim}
	\frac{1}{\sqrt{\pi k_s}}e^{i\pi/4}\exp\left(+ik_s r_s\mat{1}\right) \;, 
\label{eq:foppdef}\\
\mat{f}^{o-} &\overset{r\to\infty}{\sim} 
	\frac{1}{\sqrt{\pi k_s}}e^{i\pi/4}\exp\left(-ik_s r_s\mat{1}\right) \;, 
\label{eq:fompdef}
\end{align}
\end{subequations}
corresponding to traveling waves with fluxes of $\hbar/(\pi\mu\beta_n^{(x)})$ in the $\pm \hat{\bm{r}}$ directions, respectively. They are defined for negative energies by
\begin{subequations}
\begin{align}
\mat{f}^{o+} &\overset{r\to\infty}{\sim} 
	\frac{1}{\sqrt{\pi\kappa_s}}\exp\left(-\kappa_s r_s\mat{1}\right) \;, 
\label{eq:fopndef}\\
\mat{f}^{o-} &\overset{r\to\infty}{\sim}
	\frac{1}{\sqrt{\pi\kappa_s}}\exp\left(+\kappa_s r_s\mat{1}\right) \;. 
\label{eq:fomndef}
\end{align}
\end{subequations}
The phase factors and normalizations are chosen, similar to Ref.~\cite{Gao08a}, with the following criteria. (a) Equations~(\ref{eq:fopndef}) and (\ref{eq:fomndef}) are analytic continuations of Eqs.~(\ref{eq:foppdef}) and (\ref{eq:fompdef}) on the physical sheet, on which $(\epsilon)^{1/2}=i\kappa$ for negative energies. This allows for a consistent definition of the $S$ matrix for both positive and negative energies (with potential extension to complex energies). (b) $\mat{f}^{o+}$ and $\mat{f}^{o-}$ are both real for negative energies, while maintaining the standard definition of the $S$ matrix for positive energies. (c) They are normalized such that
\[
\mat{W}_{r_s}(\mat{f}^{o+},\mat{f}^{o-})=\frac{2}{\pi}\mat{1} \;.
\] 
Like other solutions of Eq.~(\ref{eq:aniLRss}) or (\ref{eq:aniLRsss}), $\mat{f}^{o+}$ and $\mat{f}^{o-}$ can be written as linear combinations of $\mat{f}^c$ and $\mat{g}^c$. From Eqs.~(\ref{eq:fcLRne}) and (\ref{eq:gcLRne}), it is clear that they are related, for negative energies, by the $\mat{W}^c$ matrix
\begin{align*}
\mat{f}^{c} &= \mat{f}^{o+}\mat{W}^{c}_{+f}
	+\mat{f}^{o-}\mat{W}^{c}_{-f} \;, \\
\mat{g}^{c} &= \mat{f}^{o+}\mat{W}^{c}_{+g}
	+\mat{f}^{o-}\mat{W}^{c}_{-g} \;,
\end{align*}
or in a concise matrix form as
\begin{equation}
\begin{pmatrix}
\mat{f}^c & \mat{g}^c
\end{pmatrix}
=
\begin{pmatrix}
\mat{f}^{o+} & \mat{f}^{o-}
\end{pmatrix}
\mat{W}^c
=
\begin{pmatrix}
\mat{f}^{o+} & \mat{f}^{o-}
\end{pmatrix}
\begin{pmatrix}
\mat{W}^c_{+f} & \mat{W}^c_{+g} \\
\mat{W}^c_{-f} & \mat{W}^c_{-g}
\end{pmatrix}
\;.
\label{eq:cone}
\end{equation}
The $\mat{W}^c$ matrix can thus also be understood as the matrix that relates the $(\mat{f}^c, \mat{g}^c)$ pair and the $(\mat{f}^{o+}, \mat{f}^{o-})$ pair at negative energies. For positive energies, from Eqs.~(\ref{eq:fcLRpe}) and (\ref{eq:gcLRpe}), the $(\mat{f}^c, \mat{g}^c)$ and the $(\mat{f}^{o+}, \mat{f}^{o-})$  pairs are related by
\begin{multline}
\begin{pmatrix}
\mat{f}^c & \mat{g}^c
\end{pmatrix}
= -
\begin{pmatrix}
\mat{f}^{o+} & \mat{f}^{o-}
\end{pmatrix}
\frac{1}{2}e^{-i(\mat{\ell}+\tfrac{1}{2}\mat{1})\pi/2} \\
\times
\begin{pmatrix}
(\mat{Z}^c_{cf}+i\mat{Z}^c_{sf}) & (\mat{Z}^c_{cg}+i\mat{Z}^c_{sg}) \\
e^{i\mat{\ell}\pi}(\mat{Z}^c_{cf}-i\mat{Z}^c_{sf}) & e^{i\mat{\ell}\pi}(\mat{Z}^c_{cg}-i\mat{Z}^c_{sg}) 
\end{pmatrix}
\;.
\label{eq:cope}
\end{multline}
This relation will facilitate the representations of reflection and transmission amplitudes, to be defined in the next subsection, in terms of the $\mat{Z}^c$ matrix.

The definitions of the base $(\mat{f}^c,\mat{g}^c)$ pair and the $\mat{Z}^c$ and $\mat{W}^c$ matrices are sufficient to build a $K$-matrix formulation of MQDTA using a short-range $\mat{K}^c$ matrix (see, Sec.~\ref{sec:Kmat}). The $\mat{Z}^c$ and $\mat{W}^c$ matrices are the simplest real matrices that completely describe the propagation of a wave function through the $-\mat{C}_n/r^n$ potential for positive and negative energies, respectively. They are also usually the most convenient QDT functions to compute, from which other QDT functions, such as the reflection and transmission amplitudes of the next subsection, can be calculated.

\subsubsection{Reflection and transmission amplitudes associated with the long-range potential}

The reflection and transmission amplitudes give another, physically more direct and more intuitive, way of describing the propagation of a wave function through a long-range potential, using traveling instead of standing waves. They were introduced into the QDT formulation mainly for conceptual understanding initially \cite{Gao08a}. It has since been found that an $S$-matrix formulation using those amplitudes and a short-range $S$ matrix is often the most convenient for understanding reactions and many inelastic processes, especially when the number of channels is too large for brute-force calculations \cite{Gao10b,Gao11a}. We show here how they can be defined for a generally anisotropic potential and how they relate to the $\mat{Z}^c$ and $\mat{W}^c$ matrices. 

The reflection and transmission amplitudes are closely related to the definition of another pair of reference functions $(\mat{f}^{i+}, \mat{f}^{i-})$ which define the outgoing and the incoming waves in the inner region of $r\ll \beta_n^{(x)}$, respectively. Specifically, they are defined as solutions of Eq.~(\ref{eq:aniLRss}) or (\ref{eq:aniLRsss}) that satisfy the boundary conditions
\begin{subequations}
\begin{align}
\widetilde{\mat{f}}^{i+} & := \mat{U}_{nd}^T\mat{f}^{i+} \mat{U}_{nd} \;, \\
	&\overset{r\to 0}{\sim}
	\sqrt{\frac{\mat{\beta}_{ns}}{\pi}}e^{i\pi/4}\mat{r}_s^{n/4}
	\exp\left[-i\left(\mat{y}-\tfrac{\pi}{4}\mat{1}\right)\right] \;, 
\label{eq:fipdef}\\
\widetilde{\mat{f}}^{i-} & := \mat{U}_{nd}^T\mat{f}^{i-}\mat{U}_{nd} \;, \\
	&\overset{r\to 0}{\sim} 
	\sqrt{\frac{\mat{\beta}_{ns}}{\pi}}e^{i\pi/4}\mat{r}_s^{n/4}
	\exp\left[+i\left(\mat{y}-\tfrac{\pi}{4}\mat{1}\right)\right] \;,
\label{eq:fimdef}
\end{align}
\end{subequations}
meaning that they are defined to be diagonal in the $\mat{C}_{nd}$ basis, with diagonal elements describing traveling waves, in the inner region of $r\ll \beta_n^{(x)}$, with fluxes of $\hbar/(\pi\mu\beta_n^{(x)})$ in the $\pm \hat{\bm{r}}$ directions, respectively. Note that it is the negative exponential that corresponds to the outgoing wave because $\mat{y}$ is a decreasing function of $\mat{r}_s$. The normalizations are chosen such that
\[
\mat{W}_{r_s}(\mat{f}^{i+},\mat{f}^{i-})=\frac{2}{\pi}\mat{1} \;.
\]

Comparing the definition of $(\mat{f}^{i+}, \mat{f}^{i-})$ pair with that of $(\mat{f}^{c}, \mat{g}^{c})$, it is clear that they are related by a simple unitary transformation
\begin{equation}
\begin{pmatrix}
\mat{f}^{i+} & \mat{f}^{i-} 
\end{pmatrix}
=
\begin{pmatrix}
\mat{f}^{c} & \mat{g}^{c} 
\end{pmatrix}
\frac{e^{i\pi/4}}{\sqrt{2}}
\begin{pmatrix}
	\mat{1} & \mat{1} \\
	\mat{i} & -\mat{i} 
\end{pmatrix}
\;,
\label{eq:ic}	
\end{equation}
for all energies. Substituting Eq.~(\ref{eq:cope}) into Eq.~(\ref{eq:ic}), we obtain, for $\epsilon>0$, the following relation between the $(\mat{f}^{i+}, \mat{f}^{i-})$ and $(\mat{f}^{o+}, \mat{f}^{o-})$ pairs
\begin{subequations}
\begin{align}
\begin{pmatrix}
\mat{f}^{i+} & \mat{f}^{i-} 
\end{pmatrix}
&=:
\begin{pmatrix}
\mat{f}^{o+} & \mat{f}^{o-} 
\end{pmatrix}
\mat{X}^{(oi)} \;,\\
&=:
\begin{pmatrix}
\mat{f}^{o+} & \mat{f}^{o-} 
\end{pmatrix}
\begin{pmatrix}
	\mat{X}^{(oi)}_{++} & \mat{X}^{(oi)}_{+-} \\
	\mat{X}^{(oi)}_{-+} & \mat{X}^{(oi)}_{--} 
\end{pmatrix}
\;,
\label{eq:Xoi}	
\end{align}
\end{subequations}
where we have defined a $2N_d\times 2N_d$ complex matrix $\mat{X}^{(oi)}$ with 4 $N_d\times N_d$  submatrices $\mat{X}^{(oi)}_{xy}$, given by
\begin{subequations}
\begin{align}
\mat{X}^{(oi)}_{++} =& -\frac{1}{2}e^{-i\mat{\ell}\pi/2}
	\left[(\mat{Z}^c_{cf}-\mat{Z}^c_{sg})+i(\mat{Z}^c_{sf}+\mat{Z}^c_{cg})\right] \;,
\label{eq:Xoipp}\\
\mat{X}^{(oi)}_{-+} =&  -\frac{1}{2}e^{+i\mat{\ell}\pi/2}
	\left[(\mat{Z}^c_{cf}+\mat{Z}^c_{sg})-i(\mat{Z}^c_{sf}-\mat{Z}^c_{cg})\right] \;,
\label{eq:Xoimp}\\
\mat{X}^{(oi)}_{+-} =& \left(\mat{X}^{(oi)}_{-+}\right)^* \;,\\
\mat{X}^{(oi)}_{--} =& \left(\mat{X}^{(oi)}_{++}\right)^* \;.
\end{align}
\end{subequations}
The inverse of this relation is
\begin{subequations}
\begin{align}
\begin{pmatrix}
\mat{f}^{o+} & \mat{f}^{o-} 
\end{pmatrix}
&=:
\begin{pmatrix}
\mat{f}^{i+} & \mat{f}^{i-} 
\end{pmatrix}
\mat{X}^{(io)} \;,\\
&=:
\begin{pmatrix}
\mat{f}^{i+} & \mat{f}^{i-} 
\end{pmatrix}
\begin{pmatrix}
	\mat{X}^{(io)}_{++} & \mat{X}^{(io)}_{+-} \\
	\mat{X}^{(io)}_{-+} & \mat{X}^{(io)}_{--} 
\end{pmatrix}
\;,
\label{eq:Xio}	
\end{align}
\end{subequations}
where
\begin{equation}
\mat{X}^{(io)} = \left(\mat{X}^{(oi)}\right)^{-1} \;.
\end{equation}
These relations between $(\mat{f}^{i+}, \mat{f}^{i-})$ and $(\mat{f}^{o+}, \mat{f}^{o-})$ pairs contain the reflection and transmission amplitudes, which are $N_d\times N_d$ generally non-diagonal matrices, that we now define.

For a traveling wave going outside-in, the reflection amplitude matrix $\mat{r}^{(oi)}$ and corresponding transmission amplitude matrix $\mat{t}^{(oi)}$ are defined by a solution of Eq.~(\ref{eq:aniLRss}) or (\ref{eq:aniLRsss}), $\mat{v}^{(oi)}$, with boundary conditions
\begin{align}
\mat{v}^{(oi)} \overset{r\gg \beta_n^{(x)}}{\sim}&
	\mat{f}^{o-}+\mat{f}^{o+}\mat{r}^{(oi)} \;,\nonumber\\
	\overset{r\ll  \beta_n^{(x)}}{\sim}& 
	\mat{f}^{i-}\mat{t}^{(oi)} \;.
\end{align}
Because $\mat{f}^{o\pm}$ and $\mat{f}^{i-}$ are all solutions of the same equation as the $\mat{v}^{(oi)}$ solution, the $\mat{r}^{(oi)}$ and $\mat{t}^{(oi)}$ are constant matrices, and the limit signs are equivalent to equal signs. In other words, $\mat{v}^{(oi)}=\mat{f}^{i-}\mat{t}^{(oi)}$, or more conveniently,
\begin{equation}
\mat{f}^{i-} = \mat{f}^{o+}\mat{r}^{(oi)}(\mat{t}^{(oi)})^{-1}
	+\mat{f}^{o-} (\mat{t}^{(oi)})^{-1} \;.
\label{eq:fimo}	
\end{equation}
Comparing it to Eq.~(\ref{eq:Xoi}), we obtain
\begin{multline}
\mat{t}^{(oi)} = \left(\mat{X}^{(oi)*}_{++}\right)^{-1} \\
	 = -2\left[(\mat{Z}^c_{cf}-\mat{Z}^c_{sg})-i(\mat{Z}^c_{sf}+\mat{Z}^c_{cg})\right]^{-1}
	 e^{-i\mat{\ell}\pi/2} \;,
\label{eq:toi}
\end{multline}	
\begin{align}
\mat{r}^{(oi)} = & \mat{X}^{(oi)*}_{-+}\left(\mat{X}^{(oi)*}_{++}\right)^{-1} \nonumber\\
	= & e^{-i\mat{\ell}\pi/2}\left[(\mat{Z}^c_{cf}+\mat{Z}^c_{sg})
	+i(\mat{Z}^c_{sf}-\mat{Z}^c_{cg})\right] \nonumber\\
	& \times\left[(\mat{Z}^c_{cf}-\mat{Z}^c_{sg})-i(\mat{Z}^c_{sf}+\mat{Z}^c_{cg})\right]^{-1}
	e^{-i\mat{\ell}\pi/2} \;.
\label{eq:roi}	
\end{align}	
For a traveling wave going inside-out, the reflection amplitude matrix $\mat{r}^{(io)}$ and the corresponding transmission amplitude matrix $\mat{t}^{(io)}$ are defined by a solution $\mat{v}^{(io)}$ of Eq.~(\ref{eq:aniLRss}) or (\ref{eq:aniLRsss}), with boundary conditions
\begin{align}
\mat{v}^{(io)} \overset{r_s\ll 1}{\sim}&
	\mat{f}^{i+}+\mat{f}^{i-}\mat{r}^{(io)} \;,\nonumber\\
	\overset{r_s\gg 1}{\sim}&
	\mat{f}^{o+}\mat{t}^{(io)}  \;.
\label{eq:rtiodef}	
\end{align}
It implies $\mat{v}^{(io)}=\mat{f}^{o+}\mat{t}^{(io)}$, or 
\begin{equation}
\mat{f}^{o+} =\mat{f}^{i+} (\mat{t}^{(io)})^{-1}
	+\mat{f}^{i-}\mat{r}^{(io)}(\mat{t}^{(io)})^{-1} \;.
\label{eq:fopi}	
\end{equation}
Comparing this equation with Eq.~(\ref{eq:Xio}), we obtain
\begin{align}
\mat{t}^{(io)} =& \left(\mat{X}^{(io)}_{++}\right)^{-1} \nonumber\\
	=& \mat{X}^{(oi)}_{++}-\mat{X}^{(oi)*}_{-+}
		\left(\mat{X}^{(oi)*}_{++}\right)^{-1}\mat{X}^{(oi)}_{-+} \nonumber\\
	=& -\frac{1}{2}e^{-i\mat{\ell}\pi/2}\left[(\mat{Z}^c_{cf}-\mat{Z}^c_{sg})
		+i(\mat{Z}^c_{sf}+\mat{Z}^c_{cg})\right] \nonumber\\
	& +\frac{1}{2}e^{-i\mat{\ell}\pi/2}\left[(\mat{Z}^c_{cf}+\mat{Z}^c_{sg})
	+i(\mat{Z}^c_{sf}-\mat{Z}^c_{cg})\right] \nonumber\\
	& \times\left[(\mat{Z}^c_{cf}-\mat{Z}^c_{sg})-i(\mat{Z}^c_{sf}+\mat{Z}^c_{cg})\right]^{-1} \nonumber\\
	& \times\left[(\mat{Z}^c_{cf}+\mat{Z}^c_{sg})
	-i(\mat{Z}^c_{sf}-\mat{Z}^c_{cg})\right] \;,
\label{eq:tio}	
\end{align}	
\begin{align}
\mat{r}^{(io)} =& \mat{X}^{(io)}_{-+}\left(\mat{X}^{(io)}_{++}\right)^{-1} \nonumber\\
	=& -\left(\mat{X}^{(oi)*}_{++}\right)^{-1}\mat{X}^{(oi)}_{-+} \nonumber\\
	=& -\left[(\mat{Z}^c_{cf}-\mat{Z}^c_{sg})-i(\mat{Z}^c_{sf}+\mat{Z}^c_{cg})\right]^{-1} \nonumber\\
	&\times \left[(\mat{Z}^c_{cf}+\mat{Z}^c_{sg})-i(\mat{Z}^c_{sf}-\mat{Z}^c_{cg})\right] \;.
\label{eq:rio}	
\end{align}
Computationally, it is useful to note that all transmission and reflection amplitudes are related in a simple way to two matrices, $\mat{X}^{(oi)}_{++}$ and $\mat{X}^{(oi)}_{-+}$, given in terms of the $\mat{Z}^c$ matrix by Eqs.~(\ref{eq:Xoipp}) and (\ref{eq:Xoimp}).
From the Wronskian relations for $\mat{f}^{o\pm}$ and $\mat{f}^{i\pm}$ pairs, it can be shown that the transmission and reflection amplitudes satisfy 
\begin{equation}
\mat{t}^{(oi)\dagger}\mat{t}^{(oi)}+\mat{r}^{(oi)\dagger}\mat{r}^{(oi)} = \mat{1} \;,
\end{equation}
and
\begin{equation}
\mat{t}^{(io)\dagger}\mat{t}^{(io)}+\mat{r}^{(io)\dagger}\mat{r}^{(io)} = \mat{1} \;,
\end{equation}
as representations of the conservation of probability in the propagation through the long-range potential. Here the $^\dagger$ symbol represents hermitian conjugate.

In terms of the reflection and transmission amplitudes, the relation between the $\mat{f}^{i\pm}$ pair and the $\mat{f}^{o\pm}$ pair, which contains all the information about propagation through the long-range potential for $\epsilon>0$, can be summarized as
\begin{multline}
\begin{pmatrix}
\mat{f}^{i+} & \mat{f}^{i-} 
\end{pmatrix}
=
\begin{pmatrix}
\mat{f}^{o+} & \mat{f}^{o-} 
\end{pmatrix}
\\
\times
\begin{pmatrix}
	\mat{t}^{(io)}-\mat{r}^{(oi)}\mat{t}^{(oi)})^{-1}\mat{r}^{(io)} & \mat{r}^{(oi)}\mat{t}^{(oi)})^{-1} \\
	-\mat{t}^{(oi)})^{-1}\mat{r}^{(io)} & \mat{t}^{(oi)})^{-1}
\end{pmatrix}
\;.
\label{eq:iope}	
\end{multline}
For $\epsilon<0$, the relation between the $\mat{f}^{i\pm}$ pair and the $\mat{f}^{o\pm}$ pair can be obtained by substituting Eq.~(\ref{eq:cone}) into Eq.~(\ref{eq:ic}). We have
\begin{multline}
\begin{pmatrix}
\mat{f}^{i+} & \mat{f}^{i-} 
\end{pmatrix}
= \frac{1}{\sqrt{2}}e^{i\pi/4}
\begin{pmatrix}
\mat{f}^{o+} & \mat{f}^{o-} 
\end{pmatrix}
\\
\times
\begin{pmatrix}
	\mat{W}^c_{+f}+i\mat{W}^c_{+g} & \mat{W}^c_{+f}-i\mat{W}^c_{+g} \\
	\mat{W}^c_{-f}+i\mat{W}^c_{-g} & \mat{W}^c_{-f}-i\mat{W}^c_{-g}
\end{pmatrix}
\;.
\label{eq:ione}	
\end{multline}
These relations between $\mat{f}^{i\pm}$ and $\mat{f}^{o\pm}$ pairs enable the $S$ matrix formulation of MQDTA, to be presented in Sec.~\ref{sec:Smat}.

\subsection{The $K$-matrix formulation of MQDTA}
\label{sec:Kmat}

As stated at the beginning of Sec.~\ref{sec:MQDTA}, MQDTA treats the long range of an $N_\mathrm{ch}$-channel problem by splitting it into a set of degenerate manifolds with each manifold labeled by $d$ and having $N_d$ degenerate channels. Beyond a certain radius $r_0$ where the potential has become well represented by its $-\mat{C}_n/r^n$ asymptotic form, only the coupling within the same $d$ is significant and is fully accounted for in the QDT functions defined in Sec.~\ref{sec:qdtFuncs}. All degenerate manifolds, together, make up the $\sum_d N_d = N_\mathrm{ch}$ channel problem. For every $N_d\times N_d$ QDT matrix function, call it $\mat{Y}$, defined in Sec.~\ref{sec:qdtFuncs}, there is a corresponding $N_\mathrm{ch}\times N_\mathrm{ch}$ matrix defined by
\[
(\mat{Y})_{dd} = \mat{Y}(d) \;,
\]
where $\mat{Y}(d)$ refers to the $\mat{Y}$ function of $N_d\times N_d$ dimension for the manifold $d$, and
\[
(\mat{Y})_{d'd} = \mat{0} \;,
\]
for $d'\neq d$. These block-diagonal concatenated matrices, all sharing the same $E$ but different relative energies $\epsilon_d = E-E_d$, describe long-range wave functions and their propagations of the $N_\mathrm{ch}$-channel problem. All QDT functions are thus defined in treating the full $N_\mathrm{ch}$-channel problem.

The $K$-matrix formulation of MQDTA is the computationally simplest formulation using a short-range $\mat{K}^c$ matrix and the real QDT functions $\mat{Z}^c$ and $\mat{W}^c$.
For an $N_\mathrm{ch}$ channel problem, there are $N_\mathrm{ch}$ linearly independent solutions satisfying the boundary condition at the origin, $r=0$. At sufficiently large $r\ge r_0$ where the potential has become well represented by its $-\mat{C}_n/r^n$ asymptotic form, they can be written as a linear superposition of the $(\mat{f}^c, \mat{g}^c)$ base pair,
\begin{equation}
\mat{F}^c = \mat{f}^c-\mat{g}^c \mat{K}^c \;,
\label{eq:Kcdef}
\end{equation}
which defines the short-range $\mat{K}^c$ matrix. It is real and symmetric, like the physical $K$-matrix, $\mat{K}$.

At an energy where all channels are open, all $N_\mathrm{ch}$ solution contained in $\mat{F}^c$ already satisfy the physical boundary condition at $r\to\infty$ (of being finite). Using the relation between the $(\mat{f}^c, \mat{g}^c)$ base pair and the $(\mat{s},\mat{c})$ pair as given by Eq.~(\ref{eq:cscpe}), one can easily derive from $\mat{F}^c$ a set of $N_\mathrm{ch}$ solutions that define the physical $K$-matrix, $\mat{K}$,
\begin{equation}
\mat{F} \overset{r\to\infty}{\sim} A^K (\mat{s}-\mat{c}\mat{K}) ,
\end{equation}
with the result of
\begin{equation}
\mat{K} = -( \mat{Z}^c_{cf}-\mat{Z}^c_{cg}\mat{K}^{c})
	(\mat{Z}^c_{sf} - \mat{Z}^c_{sg}\mat{K}^{c})^{-1} \;.
\label{eq:Kphyo}
\end{equation}
Here $\mat{Z}^c_{xy}$'s are examples of the QDT matrix functions, the $\mat{Y}$'s, mentioned above. They are all $N_\mathrm{ch}\times N_\mathrm{ch}$ matrices made up of a set of $N_d\times N_d$ submatrices, each associated with a threshold $E_d$, and evaluated at a relative energy of $\epsilon_{d}=E-E_d$. The $A^K$ is a normalization constant which can be chosen, for instance, as $A^K = A^K_E = (\mu\beta_n^{(x)}/\hbar^2)^{1/2}$ so that all the $N_\mathrm{ch}$ solutions in $\mat{F}$ are normalized per unit energy.

At an energy where $N_o$ channels are open, and $N_c=N_\mathrm{ch}-N_o$ channels are closed, the closed channel components in the physical solutions have to decay exponentially. These $N_c$ conditions reduce the number of linearly independent physical solutions to $N_o$, which are described by an $N_o\times N_o$ physical $K$ matrix, $\mat{K}$ (see, e.g., Ref.~\cite{Seaton83}). We obtain
\begin{equation} 
\mat{K} = -( \mat{Z}^c_{cf}-\mat{Z}^c_{cg}\mat{K}^{c}_{\mathrm{eff}})
	(\mat{Z}^c_{sf} - \mat{Z}^c_{sg}\mat{K}^{c}_{\mathrm{eff}})^{-1} \;,
\label{eq:Kphy}
\end{equation}
where
\begin{equation}
\mat{K}^c_{\mathrm{eff}} = \mat{K}^{c}_{oo}
	+\mat{K}^{c}_{oc}(\mat{\chi}^c -\mat{K}^{c}_{cc})^{-1}\mat{K}^{c}_{co} \;,
\label{eq:Kceff}
\end{equation}
in which $\mat{\chi}^c$ is an $N_c\times N_c$ real and symmetric matrix defined by
\begin{equation}
\mat{\chi}^c := (\mat{W}^c_{-g})^{-1}\mat{W}^c_{-f}\;,
\label{eq:chic}
\end{equation}
and $\mat{K}^{c}_{oo}$, $\mat{K}^{c}_{oc}$, $\mat{K}^{c}_{co}$, and $\mat{K}^{c}_{cc}$, are submatrices of $\mat{K}^{c}$ corresponding to open-open, open-closed, closed-open, and closed-closed channels, respectively. Equation~(\ref{eq:Kphy}) is the same in form as Eq.~(\ref{eq:Kphyo}), except that the $N_\mathrm{ch}\times N_\mathrm{ch}$ $\mat{K}^c$ matrix is replaced by a smaller $N_o\times N_o$ $\mat{K}^c_{\mathrm{eff}}$ that accounts for the effects of closed channels. From the physical $K$ matrix, of either Eq.~(\ref{eq:Kphyo}) or (\ref{eq:Kphy}), the physical $S$ matrix is obtained from 
\begin{equation}
\mat{S} = (\mat{1}+i\mat{K})(\mat{1}-i\mat{K})^{-1} \;, 
\label{eq:SK}
\end{equation}
from which all scattering observables can be deduced (see, e.g., \cite{Gao96,Gao10b}).

At energies where all channels are closed, the radial wave functions in all $N_\mathrm{ch}$ channels have to decay exponentially. These $N_\mathrm{ch}$ conditions can be satisfied simultaneously only at a discrete set of energies that defines the bound spectrum, which can be determined from
\begin{equation}
\det[\mat{\chi}^c(E) - \mat{K}^{c}]=0 \;,
\label{eq:bound}
\end{equation}
where $\mat{K}^c$ is the full $N_\mathrm{ch}\times N_\mathrm{ch}$ $\mat{K}^c$ matrix. For computation or physical understanding, this equation for bound spectrum can also be recasted, as appropriate, into an equivalent effective single-channel or an effective multichannel problem of smaller dimensions, as discussed in Ref.~\cite{Gao11b}.

\subsection{The $S$-matrix formulation using reflection and transmission amplitudes}
\label{sec:Smat}

MQDTA can also be formulated using reflection and transmission amplitudes and a short-range $S$-matrix $\mat{S}^c$. Here we again start with the $N_\mathrm{ch}$ linearly independent solutions satisfying the physical boundary condition at the origin. Instead of matching to the ($\mat{f}^c$, $\mat{g}^c$) pair to define the short-range $\mat{K}^c$ matrix, they are matched to the ($\mat{f}^{i+}$, $\mat{f}^{i-}$) pair to define the short-range $\mat{S}^c$ matrix, as in
\begin{equation}
\mat{F}^{S^c} = \mat{f}^{i-}+\mat{f}^{i+} \mat{S}^c \;.
\end{equation}
The $\mat{S}^c$ matrix, thus defined, has a clear physical interpretation of being the reflection amplitude by the short-range potential. It is related to the short-range $\mat{K}^c$ matrix by
\begin{equation}
\mat{S}^c = (\mat{1}+i{\mat{K}^c})(\mat{1}-i{\mat{K}^c})^{-1} \;.
\label{eq:ScKc}
\end{equation}
In other words, they are related to each other in the same manner as that between the physical $S$ and the physical $K$ matrices, as in Eq.~(\ref{eq:SK}).

From the set of $N_\mathrm{ch}$ solutions in $\mat{F}^{S^c}$, and the relations between the $(\mat{f}^{i+},\mat{f}^{i-})$ and $(\mat{f}^{o+},\mat{f}^{o-})$ pairs as given by Eqs.~(\ref{eq:iope}) and (\ref{eq:ione}) for positive $\epsilon_d$ and negative $\epsilon_d$, respectively, we can construct the $N_{o}$ solutions satisfying the scattering boundary condition at $r\to\infty$, either in a form that defines the reflection amplitude $\mat{r}^{(oi)V}$ by the full potential $\mat{V}$ 
\begin{equation}
\mat{F}^r_{oo} \overset{r\to\infty}{\sim} \mat{f}^{o-}_{oo}+\mat{f}^{o+}_{oo}\mat{r}^{(oi)V},
\end{equation}
or in a closely-related form that defines the $S$ matrix,
\begin{equation}
\mat{F}^S_{oo} \overset{r\to\infty}{\sim} \bar{\mat{f}}^{o-}_{oo}-\bar{\mat{f}}^{o+}_{oo}\mat{S},
\label{eq:Sdeffbopm}
\end{equation}
where
\begin{align}
\bar{\mat{f}}^{o+} &:= \mat{f}^{o+} e^{-i\mat{\ell}\pi/2} \;, 
\label{eq:fbopdef}\\
\bar{\mat{f}}^{o-} &:= \mat{f}^{o-} e^{+i\mat{\ell}\pi/2} \;, 
\label{eq:fbomdef}
\end{align}
and
\begin{equation}
\mat{S} = -e^{i\mat{\ell}\pi/2}\mat{r}^{(oi)V}e^{i\mat{\ell}\pi/2} \;.
\label{eq:SrV}
\end{equation}
The $\bar{\mat{f}}^{o\pm}$ pair differs from the $\mat{f}^{o\pm}$ only by phase factors. They have  asymptotic behaviors of
\begin{align}
\bar{\mat{f}}^{o+} &\overset{r\to\infty}{\sim}
	\frac{1}{\sqrt{\pi k_s}}e^{i\pi/4}\exp\left(+ik_s r_s\mat{1}-i\mat{\ell}\pi/2\right) \;, 
\label{eq:fboppasy}\\
\bar{\mat{f}}^{o-} &\overset{r\to\infty}{\sim} 
	\frac{1}{\sqrt{\pi k_s}}e^{i\pi/4}\exp\left(-ik_s r_s\mat{1}+i\mat{\ell}\pi/2\right) \;, 
\label{eq:fbompasy}
\end{align}
for $\epsilon_d>0$, and are the pair that defines the physical $S$ matrix through Eq.~(\ref{eq:Sdeffbopm}) (see, e.g., Ref.~\cite{Gao96,Gao08a}). 
Equation~(\ref{eq:SrV}) shows that the physical $S$ matrix is, up to phase factors related to partial waves, the reflection amplitude, $\mat{r}^{(oi)V}$, by the \textit{full} interaction potential, including both the long-range and the short-range portions. This understanding is helpful for the physical interpretation of the $S$ matrix and its corresponding transition amplitudes \cite{Gao08a,Gao10b}.

In the range of energies where all channels are open, we obtain, using Eq.~(\ref{eq:iope}),
\begin{widetext}
\begin{subequations}
\begin{eqnarray}
\mat{S} &=& -e^{i\mat{\ell}\pi/2}\left[\mat{r}^{(oi)}
	+\mat{t}^{(io)}\mat{S}^{c}
	(\mat{1}-\mat{r}^{(io)}\mat{S}^{c})^{-1}\mat{t}^{(oi)}\right]
	e^{i\mat{\ell}\pi/2} \;,
\label{eq:SmqdtSo} \\
&=& -e^{i\mat{\ell}\pi/2}\left\{\mat{r}^{(oi)}
	+\mat{t}^{(io)}\mat{S}^{c}
	\left[\sum_{m=0}^{\infty}(\mat{r}^{(io)}\mat{S}^{c})^m
	\right]\mat{t}^{(oi)}\right\}e^{i\mat{\ell}\pi/2} \;,
\label{eq:SmqdtSoe}
\end{eqnarray}
\end{subequations}
\end{widetext}
where the reflection and transmission amplitudes are all $N_\mathrm{ch}\times N_\mathrm{ch}$ matrices made up of a set of $N_d\times N_d$ submatrices, each associated with a threshold $E_d$, and evaluated at a relative energy of $\epsilon_{d}=E-E_d$.

In the range of energies where $N_o$ channels are open, and $N_c=N_\mathrm{ch}-N_o$ channels are closed, the number of linearly independent physical scattering solutions is reduced to $N_o$. We obtain, for the $N_o\times N_o$ physical $S$ matrix,
\begin{widetext}
\begin{subequations}
\begin{eqnarray}
\mat{S} &=& -e^{i\mat{\ell}\pi/2}\left[\mat{r}^{(oi)}_{oo}
	+\mat{t}^{(io)}_{oo}\mat{S}^{c}_{\mathrm{eff}}
	(\mat{1}-\mat{r}^{(io)}_{oo}\mat{S}^{c}_{\mathrm{eff}})^{-1}\mat{t}^{(oi)}_{oo}\right]
	e^{i\mat{\ell}\pi/2} \;,
\label{eq:SmqdtS} \\
&=& -e^{i\mat{\ell}\pi/2}\left\{\mat{r}^{(oi)}_{oo}
	+\mat{t}^{(io)}_{oo}\mat{S}^{c}_{\mathrm{eff}}
	\left[\sum_{m=0}^{\infty}(\mat{r}^{(io)}_{oo}\mat{S}^{c}_{\mathrm{eff}})^m
	\right]\mat{t}^{(oi)}_{oo}\right\}e^{i\mat{\ell}\pi/2} \;,
\label{eq:SmqdtSe}
\end{eqnarray}
\end{subequations}
\end{widetext}
using Eqs.~(\ref{eq:iope}) and (\ref{eq:ione}). Here the reflection and transmission amplitudes are  $N_o\times N_o$ matrices defined for the combination of all open channels. The $\mat{S}^c_{\mathrm{eff}}$ is an  $N_o\times N_o$ effective short-range $S$ matrix that encapsulates all effects of closed channels,
\begin{equation}
\mat{S}^c_{\mathrm{eff}} = \mat{S}^{c}_{oo}
	+\mat{S}^{c}_{oc}(\mat{\chi}^s -\mat{S}^{c}_{cc})^{-1}\mat{S}^{c}_{co} \;,
\label{eq:Sceff}
\end{equation}
in which $\mat{S}^{c}_{oo}$, $\mat{S}^{c}_{oc}$, $\mat{S}^{c}_{co}$, and $\mat{S}^{c}_{cc}$, are submatrices of $\mat{S}^{c}$ corresponding to open-open, open-closed, closed-open, and closed-closed channels, respectively, and
\begin{align}
\mat{\chi}^s &= (\mat{1}+i{\mat{\chi}^c})(\mat{1}-i{\mat{\chi}^c})^{-1}\nonumber\\
	&= (\mat{1}-i{\mat{\chi}^c})^{-1}(\mat{1}+i{\mat{\chi}^c}) \;,
\label{eq:chis}
\end{align}
is a unitary matrix that relates to $\mat{\chi}^c$ in the same way as an $S$ to a $K$ matrix in general.

Equations~(\ref{eq:SmqdtSo}) and (\ref{eq:SmqdtS}) give the physical $S$ matrix, $\mat{S}$, in terms of reflection and transmission amplitudes associated with the long-range potential and a short-range $S$ matrix, $\mat{S}^c$, or an effective short-range $S$ matrix, $\mat{S}^c_{\mathrm{eff}}$, which, similar to $\mat{S}^c$, has the physical meaning of being an effective reflection amplitude by the inner potential. These representations of the $S$ matrix have clear physical interpretations similar to those discussed for the isotropic cases in Refs.~\cite{Gao08a,Gao10b}, and expressed here in their expanded forms in Eqs.~(\ref{eq:SmqdtSoe}) and (\ref{eq:SmqdtSe}). They show explicitly  the $S$ matrix or the corresponding transition amplitude being 
made of coherent contributions from multiple paths. In particular, the $m$-th term in the expansion, in either Eq.~(\ref{eq:SmqdtSoe}) or Eq.~(\ref{eq:SmqdtSe}), corresponds to a contribution from a path in which the particles are reflected $m+1$ times by the inner potential, each time represented by one power of $\mat{S}^{c}$ or $\mat{S}^{c}_{\mathrm{eff}}$.

In the energy region where all channels are closed, the bound spectrum is still most conveniently determined by Eq.~(\ref{eq:bound}). It is however of interest mathematically that the bound spectrum can also be determined from
\begin{equation}
\det[\mat{\chi}^s(E) - \mat{S}^{c}]=0 \;.
\label{eq:boundS}
\end{equation}
This option means that the formulation using $\mat{S}^c$ is mathematically complete. We note that while the physical $S$ matrix is defined only for open channels, the short-range $S$ matrix, $\mat{S}^c$, is well defined for all channels.

\section{Discussions}
\label{sec:discussions}

\subsection{General}
\label{sec:GD}

It is gratifying that with some trivial changes in notation (see Appendix~\ref{sec:notations} for a summary), the equations of MQDTA are the same in form as those for isotropic potentials \cite{Gao05a,Gao10b}. The key difference is that the QDT function matrices, such as $\mat{Z}^c_{xy}$ and $\mat{\chi}^c$, are no longer fully diagonal as they are for isotropic long-range potentials \cite{Gao05a,Gao10b}. They are instead block-diagonal, made of blocks of $N_d\times N_d$ dimensions, one for each degenerate manifold. The functions for each block come from solutions of the QDT equation for a generally anisotropic potential, Eq.~(\ref{eq:aniLRss}) or (\ref{eq:aniLRsss}), which itself is generally a multichannel coupled equation. While this coupled equation is more difficult than the corresponding single-channel equation, it has a well-defined standard form that should promote further mathematical investigations, both in terms of possible analytic solutions in the spirits of Refs.~\cite{Gao98a,Gao99a,Gao99b,Gao13c}, and in terms of more specialized numerical methods built upon existing techniques (see, e.g., Refs.~\cite{Gordon69,Johnson78,Manolopoulos86,Karman14}). The broad range of systems that each solution would help to characterize should provide considerable motivation. 

The two formulations presented, the $K$ matrix formulation of Sec.~\ref{sec:Kmat} and the $S$ matrix formulation of Sec.~\ref{sec:Smat}, are mathematically equivalent, but have distinct applications and approximate forms. The $S$ matrix formulation is most convenient for the visualization of complex processes and for making approximations that take advantage of a large number of channels. It will be the cornerstone of a general MQDTA theory for chemical reactions \cite{DG20c}, that goes far beyond the earlier universal theories for exoergic processes \cite{Gao10b,Gao11a}. The MQDTA theory will provide new insights even on old and fundamental topics such as the origin of the Wigner threshold behaviors \cite{Wigner48}.

The $K$ matrix formulation, being the simplest for computation, is also the most convenient for understanding resonances of all types including shape resonances, Feshbach resonances \cite{Chin10}, and diffraction resonances \cite{Gao10a,Gao13c,LYG14}. A deeper understanding of resonances (see, e.g., Refs.~\cite{Gao09a,Gao11b}) is crucial for constructing proper effective potentials in the energy region of resonances, in particular the region of Feshbach resonances corresponding to partially-excited internal degrees of freedom. Few-body physics and many-body physics in such regimes, with the exception of cases around a broad $s$ wave Feshbach resonance \cite{Leggett06,GPS08,Greene17}, are far from being understood (see, e.g. Refs.~\cite{FWG03,Hazlett12,LLLG18a,Chapurin19}), and are likely to remain the focal point in their respective fields. An effective potential based on MQDTA will bring a new perspective to this topic. 

Our formulation of MQDTA allows for a systematic progression towards shorter length scales \cite{Gao16a,Hood20a}. One can start with a single-scale theory, and expand it to shorter length scales through multiscale theories as needed. We emphasize that all formulations are exact if one uses exact short-range parameters. The difference is the degree to which their corresponding short-range parameters are independent of the energy and the partial waves. The choice thus depends on the range of energy one would like to cover and the degree of accuracy one would like to achieve if one is to make an approximation of constant or near-constant parameters \cite{Gao16a,Hood20a}. While we have specialized to a purely attractive single-scale long-range potential of the form $-\mat{C}_n/r^n$, partly to make it easier to understand through a more direct comparison with previous isotropic formulations \cite{Gao05a,Gao08a,Gao10b}, it should be clear that a generalization to the multiscale long-range potential of Eq.~(\ref{eq:LRVms}) is possible and conceptually straightforward. We mention two cases where multiscale theories are especially desirable. One is the case when there are competing long-range interactions of comparable length scales, such as a van der Waals $1/r^6$ potential and a strong magnetic dipole-dipole $1/r^3$ potential for highly-magnetic atoms \cite{Kotochigova14}. The other is in ion-ion or ion-polar types of atom-atom, atom-molecule, or molecule-molecule interactions. For such systems, the monopole-monopole and the monopole-dipole terms have to be included if present. At the same time, it is highly beneficial to include up to the $1/r^4$ (mostly polarization) potential if we are to explicitly take advantage of the partial-wave-insensitive nature of the short-range interaction. More aspects of multiscale theories will be explored in future studies, including the question of how to best connect theories at different scales so that extra parameters are used only when necessary and are never wasted. 

MQDTA can also accommodate cases of purely repulsive potentials \cite{Gao99a} and mixed potentials, though the purely repulsive cases are generally less interesting when considered independently \cite{Gao17a}. Further developments in these areas await sufficiently interesting applications.

\subsection{Reactions}
\label{sec:reaction}

We have avoided explicit discussion of reactions (rearrangement collisions) \cite{HS06,Hazra14b,Bala16}  intentionally. First, there are many cases of interactions, especially atom-atom interactions, where rearrangement is not possible, and inclusion of such discussion would unnecessarily add to the burden of understanding and the complexity of notation. Second, it is interesting to note explicitly how the same theory, including all essential equations, can be used for rearrangement collisions with only a slight expansion of the meaning of the index $d$. Specifically, all MQDTA equations are applicable provided that a degenerate manifold $d$ is understood as potentially belonging to a different arrangement, in which case it may have a different arrangement-specific $\mu$, a different arrangement-specific long-range potential, and different arrangement-specific channel energies. When reaction is the focus, one can label different arrangements more explicitly by, e.g., expanding the index $d$ to $\gamma d$, in which $\gamma$ is used to label different arrangements. This apparent similarity between reactive and nonreactive theories should not, however, obscure the many unique characteristics of reactions, especially chemical reactions, which will be discussed in more detail elsewhere  \cite{DG20c}. 

\subsection{External fields}
\label{sec:extFields}

Interaction in external fields \cite{Bala16} is beyond the scope of this work, if the fields are more than probes. The anisotropies treated in this work are those intrinsic to atomic and molecular interactions. They are not those due to external fields, though there can be mathematical similarities in their treatments. 

Interaction in or with a photon or laser field is different if the photon or laser field is only a probe. Thus combining the MQDTA treatments here for two different electronic states can lead directly to a theory of photoassociation \cite{Jones06}, for instance. Not being able to treat the $S+P$ electronic state in the same manner that we had treated the $S+S$ ground electronic state \cite{Gao05a}, thus not having a full MQDT for photoassociation, was one of the early disappointments with the MQDT for isotropic potentials, thus also one of the early motivations for MQDTA.

\subsection{Interactions with an electron}

The MQDTA presented here applies equally well if one of the particles is an electron, as in electron-atom \cite{WG80,Fabrikant86,Buckman94} or electron-molecule interactions \cite{Mittleman65,Clark79,Lane80,Curik06,Tennyson10,Fabrikant16}. It also applies to electron-ion \cite{Seaton83,Aymar96,Burke11} or electron-molecular-ion interactions \cite{Fano70,CF72,Greene85,Jungen96,JR98,DKG09} if we put it in a broader context of a multiscale theory that includes the monopole related potential terms, as needed. The main difference of an electron theory is that it is \textit{not} partial-wave-insensitive, as the partial-wave term $\hat{\bm{\ell}}^2/2\mu r^2$ in Eq.~(\ref{eq:Ham}) is, for an electron, comparable to or even dominant over the potential energy term $\hat{V}$ at the short range. This characteristic does not, however, leads to any difficulty, because for an electron the room temperature range of energies corresponds to the ``ultracold'' regime where the $s$ wave dominates, and even at an energy of the order of 1 eV, there are only a few contributing partial waves. The promising prospects of multiscale theories can already be seen in the 2-scale ($-C_1/r-C_4/r^4$) isotropic theory of Ref.~\cite{Gao16a}, which, with the generalization to anisotropic cores enabled by the MQDTA here, can now be extended to  other atomic and molecular species.

\subsection{Short-range parameters}
\label{sec:FT}

In all cases where the short-range calculation can be carried out, the short-range $\mat{K}^c$ matirx is obtained by matching the short-range solution onto the QDT functions, as in Eq.~(\ref{eq:Kcdef}). There are different options of numerical methods for the short-range propagation, such as the R-matrix method for electron-related interactions \cite{Aymar96,Tennyson10,Burke11}, CC methods for atom-atom interactions \cite{Tiesinga93,Samuelis00,vanKempen02,Pires14} and nonreactive atom-molecule and/or molecule-molecule interactions \cite{Toennies69,Green75,Tscherbul10,Hazra14a}, and methods based on hyperspherical coordinates \cite{Delves58,Delves60} for atom-molecule interactions and other few-atom problems \cite{PP87,SCM00,TK08,Hazra14b,Makrides15,Croft17a,Greene17}.

For complex interactions, an important question is how to most effectively parameterize the short-range $\mat{K}^c$ or $\mat{S}^c$ matrix. From the rigidity of the short-range wave function and the structure of the short-range equation, we know that the $\mat{K}^c$ or $\mat{S}^c$ matrix can be parameterized with a few energy- and partial-wave-insensitive parameters. Specifically, we expect, to the lowest order, approximately one parameter per potential energy curve (PEC) and a few parameters per potential energy surface (PES), with allowance for a few more coupling constants if there are multiple curves and/or surfaces that cross or interact strongly in the short range. For atom-atom interactions, this efficient parameterization is accomplished through a frame transformation \cite{Gao96,Gao05a}, and has been fully demonstrated for alkali-alkali interactions in the context of isotropic MQDT \cite{Gao05a,Hanna09,LYG14,MG14,LG15,Cui17b,LiuXP18,Cui18b,Hood20a,Wang20}. In MQDTA, the more general frame transformations of Ref.~\cite{Gao96} enable similar treatments of other atomic species. 

For atom-molecule and molecule-molecule interactions, the efficient representation of $\mat{K}^c$ in terms of a few parameters is much less established. The essence is still the frame transformation concept, pioneered by Fano and coworkers \cite{Fano70,RF71,CF72,Gao96,JR98,Gao05a}, that relates the basis which best describes the short-range interaction, the condensation channels, and the QDT basis which best describes the long-range region. These transformations, which are different for atom-molecule and molecule-molecule types, are under investigation and will be further discussed as we specialize to each class of systems in specific applications. We only note here that the frame-transformation-based parameterization of $\mat{K}^c$ or $\mat{S}^c$ is of growing significance both because of the increasing complexity of the systems of interest and because of the development of multiscale theories to shorter length scales.  In a multiscale theory, a well-formulated frame transformation can become more accurate at shorter length scales, and there can be a limit in which it becomes exact.

\section{Conclusions}
\label{sec:conclusions}

In conclusion, we have presented MQDTA -- an MQDT for generally anisotropic long-range potentials, providing  a framework for a systematic understanding of atomic and molecular interactions of all types, and a framework for the development of the corresponding effective potentials at different length scales. By making use of the $1/r$-property of the electromagnetic interaction and the rigidity property of a wave equation, MQDTA automatically and systematically factors out the rapid energy- and partial-wave-dependences due to the long-range potential, and leaves what to be determined to a few energy- and partial-wave-insensitive short-range parameters.

Through future efforts of implementation for specific classes of systems, the theory will have the effect of making the Periodical Table quantitative. Just like all alkali-alkali interactions can be described quantitatively with the same set of a few parameters \cite{Gao05a,Hanna09,LYG14}, similar MQDTA descriptions can now be developed for interactions between other groups with anisotropic potentials such as the group I-group IV interactions, which include, e.g., hydrogen-carbon interaction. The theory has further laid the foundation for similar systematic understandings of atom-molecule and molecule-molecule interactions.

An important characteristic of the theory is that it can function as an effective theory for interactions of complex systems, for which \textit{ab initio} calculations are either not possible or not sufficiently accurate. For such systems, the theory provides an optional description using a few parameters that are fully determined experimentally, similar to what have been demonstrated for atom-atom interactions \cite{Gao98b,Gao01,Hood20a}. 

As a part of understanding atom-molecule and molecule-molecule interactions, the framework provides a general theory of bimolecular chemical reactions that is far more complete than the earlier quantum Langevin (QL) models for exoergic processes \cite{Gao10b,Gao11a}, which were based on MQDT for isotropic potentials. This topic will be further addressed in a separate publication \cite{DG20c}.

The same framework for understanding interactions will provide a framework for constructing effective potentials and corresponding effective theories for $N$-body quantum systems of interacting atoms and/or molecules. Such theories can be tested, and have their parameters determined, using cold atoms and molecules. Once properly built, the same theories can be used to treat quantum effects at much higher temperatures including room temperatures. 

Finally, with proper adaptations, the theory can be generalized to atomic and molecular interactions with either a surface or a large molecule, retaining similar characteristics. This can eventually lead to quantum theories of $N$-body systems in which one or a few of the bodies can be a large molecule, a mesoscopic particle, or a macroscopic object. Understanding such systems at a quantum level will bring us much closer to a quantitative understanding of functionalities, such as the stability, growth, and catalytic properties of molecules, the electron and ion transport properties through different media, and the dynamics of phase transitions.

\begin{acknowledgments}
This work is supported by NSF under grants PHY-1607256 and PHY-1912489. I would like to thank Dr. Ningyi Du for discussions and suggestions, and for computational support that has helped to verify many of the concepts presented here. I thank my late Ph.D. adviser, Prof. Anthony F. Starace, for introducing me to the humble field of AMO physics, in which the spirit of science may have been a bit easier to uphold. I thank Prof. Alexander Dalgarno and Prof. Michael Cavagnero. Their timely encouragements and their appreciation of my work have been crucial to keep me afloat in physics. Finally, I would like to thank my students Ming Li and Constantinos Makrides, for staying with me through the difficult times when we had no grant support, and for their contributions in getting us to this point of clearer possibilities.
\end{acknowledgments}

\appendix

\section{Comments on the representation of anisotropic long-range potentials}
\label{sec:aniLRV}

Long-range atomic and molecular potentials of all types have been widely studied, and fairly well understood (see, e.g., Refs.~\cite{Pack76,Leavitt80,Derevianko99,Derevianko01,Groenenboom07,Stone13}). While the details can vary, they are all of the form of Eq.~(\ref{eq:LRVop}). We are making a broad suggestion that each $1/r^m$ term in Eq.~(\ref{eq:LRVop}), namely each $\hat{C}_m$ operator, be described by a $C^{(x)}_m$ parameter measuring its overall strength and a few additional dimensionless ``anisotropy parameters'', if necessary. We emphasize, however, that MQDTA does not in any way require or depend on such a parameterization. Even the definition of the strength parameter $C^{(x)}_m$ is not rigid. One can in principle define it differently provided that the definition gives the correct relative strengths and relative length scales for different terms in the potential.

As a concrete example, consider an important class of atom-molecule interactions, the type of $S+\Sigma$, for which the leading long-range interaction is of the form (see, e.g., \cite{Pack76})
\begin{equation}
V(r,\Theta) \sim -\frac{C_6(\Theta)}{r^6}-\frac{C_7(\Theta)}{r^7}-\frac{C_8(\Theta)}{r^8} \;,
\label{eq:VLR}
\end{equation}
where $\Theta$ is the angle between $\hat{\bm{r}}$ and the molecule axis, and
\begin{subequations}
\begin{align}
C_6(\Theta) &= C^{(0)}_6+C^{(2)}_6 P_2(\cos\Theta) \nonumber\\
	&= C^{(0)}_6[1+q_{6,2} P_2(\cos\Theta)] \;,\\
C_7(\Theta) &= C^{(1)}_7 P_1(\cos\Theta)+C^{(3)}_7 P_3(\cos\Theta) \nonumber\\
	&= C^{(1)}_7[P_1(\cos\Theta)+q_{7,3} P_3(\cos\Theta)] \;,\\
C_8(\Theta) &= C^{(0)}_8+C^{(2)}_8 P_2(\cos\Theta)+C^{(4)}_8 P_4(\cos\Theta) \nonumber\\
	&= C^{(0)}_8[1+q_{8,2} P_2(\cos\Theta)+q_{8,4} P_4(\cos\Theta)] \;,
\end{align}
\end{subequations}
in which we have defined $q_{6,2} := C_6^{(2)}/C_6^{(0)}$, $q_{7,3} := C_7^{(3)}/C_7^{(1)}$, $q_{8,2} := C_8^{(2)}/C_8^{(0)}$, and $q_{8,4} := C_8^{(4)}/C_8^{(0)}$. Thus the $1/r^6$ term can be characterized by 2 parameters: $C^{(0)}_6$ for the overall strength and $q_{6,2}$ for the anisotropy. The $1/r^7$ term can be characterized similarly by $C^{(1)}_7$ and $q_{7,3}$, and the $1/r^8$ term by 3 parameters: $C^{(0)}_8$, $q_{8,2}$, and $q_{8,4}$. With such a convention, in a single-scale MQDTA for the $1/r^6$ term only, the QDT matrix functions, such as the $\mat{\chi}^c$, are then functions of a scaled energy $\epsilon_s$ and $q_{6,2}$, with the energy being scaled by the isotropic $C_6$, namely $C_6^{(0)}$. 

While this form of parameterization of an anisotropic atom-molecule potential has been standard for some time (see, e.g., Ref.~\cite{Toennies69}), few conventions exist for other types of anisotropic potentials. We expect more conventions will be established as more types and systems are investigated.

\section{Comments on notations}
\label{sec:notations}

We briefly comment on changes of notation from the earlier related works \cite{Gao05a,Gao08a}, and the reason behind the changes.

The main changes of notation are that the order of the subscripts for the $\mat{Z}^c$ and $\mat{W}^c$ submatrices are reversed, such as $\mat{Z}^c_{fs}$ being changed to $\mat{Z}^c_{sf}$. This is to make the notation more consistent with the order of the matrix multiplication as in, e.g., Eq.~(\ref{eq:cscpe}). The ordering did not matter in isotropic cases where the corresponding quantity was either a pure number, in the case of a single channel \cite{Gao08a}, or a diagonal matrix \cite{Gao05a}.

The matrices $\mat{X}^{(oi)}$ and $\mat{X}^{(io)}$ correspond to $U^{(oi)}$ and $U^{(io)}$ of Ref.~\cite{Gao08a}, respectively. This change is partly to emphasize that $\mat{X}$ is generally \textit{not} unitary, as the letter $U$ might have wrongly suggested. The letter $X$, interpreted as ``crossing'', also better represents the underlying physical meaning of $\mat{X}$ as describing the propagation through a long-range potential.

For interactions in the absence of external fields, the total angular momentum $F_t$ and the total parity $\sigma_t$ are conserved, and we could have chosen to label the solutions and scattering matrices, such as $\mat{S}$ with one or both of them, e.g., as $\mat{S}^{(F_t)}$ or $\mat{S}^{(F_t,\sigma_t)}$. We left them out both for simplicity, and for the fact that many aspects of the theory will remain the same for interactions in external fields, in which case the conserved quantity will change, e.g., to a projection of the total angular momentum $M_t$. It is thus better to leave out detailed labels in a general formalism. They can be added, as needed, in specific applications.


\begin{thebibliography}{125}
\bibitem [{fun()}]{functionality}%
  \BibitemOpen
  \href@noop {} {}\bibinfo {note} {It is better to leave the definition of
  ``functionality'' vague as it prompts more thinking about what it is. It can
  be defined to refer to any property of a system that is not structural, or
  more specifically as a property when parts of a system are moving or far from
  the equilibrium, such as any property having to do with scattering or
  transport, reaction, growth, or breakup of a system.}\BibitemShut {Stop}%
\bibitem [{rig()}]{rigidity}%
  \BibitemOpen
  \href@noop {} {}\bibinfo {note} {There exists a mathematical interpretation for
  this property to be a fundamental property of a wave equation. This
  interpretation is not essential and will be presented elsewhere.}\BibitemShut
  {Stop}%
\bibitem [{pos()}]{positiveDefinite}%
  \BibitemOpen
  \href@noop {} {}\bibinfo {note} {$\mat{C}_n$ being positive definite implies
  that $(\mat{C}_n)_{dd}$ being positive definite for all $d$.}\BibitemShut
  {Stop}%
\end{thebibliography}
\end{document}